\documentclass[12pt]{article}
\newcommand{\be}{\begin{equation}}
\newcommand{\ee}{\end{equation}}

\newcommand{\bi}{\begin{itemize}}
\newcommand{\ei}{\end{itemize}}
\newcommand{\bea}{\begin{eqnarray}}
\newcommand{\eea}{\end{eqnarray}}
\newcommand{\ba}{\begin{array}}
\newcommand{\ea}{\end{array}}

\usepackage[left=2.50cm, right=2.50cm, top=2.50cm, bottom=2.50cm]{geometry}
\usepackage[utf8]{inputenc}
\usepackage{cancel}
\usepackage[english]{babel}
\usepackage{physics}
\usepackage{hyperref}
\usepackage{amsthm}
\usepackage{mathrsfs}
\usepackage{mathtools}
\usepackage{graphicx}
\usepackage{changepage}
\usepackage{amssymb,amsmath}
\usepackage{verbatim}
\usepackage{empheq}
\usepackage{xcolor}
\usepackage{tikz}
\usepackage{float}
\usepackage{multirow}
\usepackage{multicol}
\usepackage{centernot}
\usepackage{tensor}

\usepackage[hang,flushmargin]{footmisc}
\usepackage{cite}

\numberwithin{equation}{section}
\hypersetup{hidelinks}

\newlength{\bibitemsep}
\newlength{\bibparskip}
\let\oldthebibliography\thebibliography
\renewcommand\thebibliography[1]{%
  \oldthebibliography{1}%
  \setlength{\parskip}{\bibitemsep}%
  \setlength{\itemsep}{\bibparskip}%
}
\begin{document}

%%%%%%%%%%%%% Title %%%%%%%%%%%%%%%%%%%%%%%%%%

\begingroup
\linespread{1.0}\selectfont

\par\bigskip
\Large\noindent
{\bf 
Edge dynamics of covariant higher-rank Chern-Simons theory of fractons} 

%\today   

\bigskip\par\rm\normalsize

\hrule
\vspace{0.7cm}
\large\noindent
{\bf Erica Bertolini$^{1,a}$},
{\bf Matteo Carrega$^{2,b}$},
{\bf Nicola Maggiore$^{3,4,c}$},\\
{\bf Daniel Sacco Shaikh$^{5,d}$},
{\bf Maria Florencia Semino$^{3,e}$}

\par\vspace{0.50cm}\small

\noindent$^1$ School of Theoretical Physics, Dublin Institute for Advanced Studies, 10 Burlington Road, Dublin 4, Ireland.

\noindent$^2$ CNR-SPIN, Via Dodecaneso 33, 16146 Genova, Italy.

\noindent$^3$ Dipartimento di Fisica, Universit\`a di Genova, Via Dodecaneso 33, I-16146 Genova, Italy.

\noindent$^4$ Istituto Nazionale di Fisica Nucleare -- Sezione di Genova, Via Dodecaneso 33, I-16146 Genova, Italy.

\noindent$^5$ Division of Theoretical Physics, Ru{\dj}er Bo\v{s}kovi\'c Institute, Bijenicka cesta 54, 10000 Zagreb, Croatia.

\vspace{0.55cm}

\noindent{\tt Abstract}\\
We study the covariant higher-rank Chern-Simons theory of fractons in $(2+1)$ dimensions in the presence of a planar boundary. The differential generator associated with the scalar gauge symmetry singles out a single scalar Ward combination of the functional field equations. Its zeroth and first normal moments give two integrated boundary identities. The zeroth moment yields a traceless rank-two current obeying a fractonic double-divergence conservation
law and a higher-rank Kac-Moody algebra. The first moment makes the normal-tangential field $w_i=h_{2i}$ explicit and yields an ordinarily conserved vector current with a standard $U(1)$ Kac-Moody algebra. Functional differentiation also produces a mixed equal-time bulk-boundary commutator between the spatial components $w_1$ and $h_{11}$: it is non-vanishing at finite normal distance and vanishes in the intrinsic boundary limit. The equal-time commutators of the boundary current with the spatial components of the gauge-invariant field strength vanish, so microcausality is satisfied; when the boundary parameters are such that the two boundary sectors decouple, the corresponding retarded response vanishes at all times. Locally, the boundary constraints are solved by $w_i=\partial_i\chi$ and $h_{ij}=\partial_i\partial_j\xi+\eta_{ij}\phi$. When the boundary parameters are such that the two sectors decouple, the induced
action is the direct sum of a chiral boson in the $w_i$ sector and a higher-rank sector in which $\phi$ is auxiliary and $\psi=\partial_1\xi$ is the propagating field. The latter has a non-chiral Luttinger-liquid description on its positive-energy branch and admits a chiral reduction. The bulk boundary conditions determine the velocity and positivity branch of the $w_i$ mode, while the compatibility analysis in the $h_{ij}$ sector selects the positive-energy chiral branch. On the matched positive-energy branch the two chiral modes co-propagate, with orientation opposite to the sign of $\kappa$. We also identify the extrinsic-curvature terms through which non-planar geometry enters the projected gauge transformations and can consequently modify the boundary conditions and edge dynamics.

\vspace{0.45cm}

\noindent{\tt Keywords:}\\
Higher-rank gauge theory; Chern-Simons theory; fractons; boundary field theory; Kac-Moody algebra; Luttinger liquid; bulk-boundary correspondence.

\vspace{0.55cm}
\hrule
\vspace{0.25cm}
{\footnotesize\tt\raggedright
\noindent E-mail:
$^a$ebertolini@stp.dias.ie,\quad\allowbreak
$^b$matteo.carrega@spin.cnr.it,\quad\allowbreak
$^c$nicola.maggiore@ge.infn.it,\quad\allowbreak
$^d$daniel.saccoshaikh@irb.hr,\quad\allowbreak
$^e$m.florencia.semino@gmail.com.\par}

\endgroup
\linespread{1.2}\selectfont

%\tableofcontents

%**********************************************
\section{Introduction}
%**********************************************

Fracton systems have attracted considerable attention because they realise phases of matter in which the mobility of excitations is strongly constrained
\cite{Chamon:2004lew, Haah:2011drr, Bravyi:2011faf, Vijay:2015mka, Vijay:2016phm, Nandkishore:2018sel, Pretko:2020cko}.
The characteristic feature is that isolated charges cannot move freely, while suitable bound states, such as dipoles, may still propagate. In continuum field theory this phenomenon is naturally described by higher-rank gauge theories, where generalised Gauss laws imply the conservation not only of charge, but also of higher multipole moments~\cite{Pretko:2016kxt, Pretko:2016lgv, Ma:2018nhd}. In the scalar-charge theory, for instance, the conservation of dipole moment is directly responsible for the immobility of isolated fractons and for the mobility of dipolar excitations \cite{Pretko:2016kxt, Pretko:2016lgv}. 
This has led to a broad class of effective field-theoretical descriptions of fractonic matter, including higher-rank electromagnetism~\cite{Pretko:2016kxt, Pretko:2016lgv, Bertolini:2022ijb}, elasticity-inspired formulations
\cite{Pretko:2017kvd}, vector gauge theories~\cite{Radzihovsky:2019jdo},
hydrodynamic descriptions~\cite{Gromov:2020yoc, Glorioso:2021bif}, exotic subsystem and foliated descriptions
\cite{Vijay:2015mka, Vijay:2016phm, Seiberg:2020bhn, Shirley:2018vtc, Slagle:2018swq},
and field theories with non-trivial boundary dynamics \cite{Prem:2017kxc, Parasar:2025ngk}.  
In (2+1) dimensions, a particularly direct motivation is provided by the fracton-elasticity duality \cite{Pretko:2017kvd}, which maps disclinations and dislocations of an elastic medium to fractons and dipoles, respectively. Understanding the boundary dynamics of fractonic theories in three spacetime dimensions therefore bears direct relevance for the edge physics of elastic or crystalline media and for possible realisations in condensed-matter systems.

A particularly interesting direction is the construction of Chern-Simons-like theories for higher-rank gauge fields. Ordinary Chern-Simons theory plays a central role in the effective description of topological phases and quantum Hall systems \cite{Witten:1988hf, Elitzur:1989nr, Dunne:1998qy, Tong:2016kpv}. Its boundary theory is not an accessory feature, but one of the clearest physical manifestations of the bulk topological structure: the presence of a boundary induces a non-trivial Kac-Moody current algebra and lower-dimensional degrees of freedom \cite{Witten:1988hf, Elitzur:1989nr}. A well-established consequence is the connection between the boundary Kac-Moody algebra and the chiral edge modes of fractional quantum Hall states, which are described by a chiral Luttinger liquid \cite{Wen:1990se, Maggiore:2017vjf, Maggiore:2018bxr, Bertolini:2021iku}. It is therefore natural to ask whether an analogous mechanism occurs for higher-rank Chern-Simons theories, and whether their boundaries support an algebraic and dynamical structure characteristic of fractonic phases. An analogous mechanism linking a boundary Kac-Moody algebra to lower-dimensional edge dynamics was found in the boundary analysis of the four-dimensional theory of a symmetric tensor field, relevant both to fracton gauge theory and to linearized gravity \cite{Bertolini:2023juh, Bertolini:2023sqa, Bertolini:2023wie}.

Higher-rank Chern-Simons and BF theories have already been considered in non-covariant fractonic settings \cite{You:2019xcr}. Earlier related constructions appeared in the study of higher-multipole contributions to the quantum Hall effect \cite{Cappelli:2015ocj} and in the context of topological Chern-Simons elasticity with fracton content \cite{Gromov:2017vir}. Moreover, the multipole structure of such theories has been linked to higher-order topological insulators and fractonic phases \cite{You:2019bvu}. Within the same covariant three-dimensional approach to fractonic gauge theories \cite{Bertolini:2025qcy,Bertolini:2025jul}, the higher-rank Chern-Simons theory which we study here was introduced in Ref.~\cite{Bertolini:2024yur}. In that work the theory was analysed in the absence of a boundary and its fractonic character was made explicit. In particular, once the theory is coupled to matter, its equations of motion can be interpreted in terms of higher-rank electric and magnetic-like fields, thus reproducing Hall-like dipole transport and a vectorial flux-attachment relation of a higher-rank planar electromagnetism. The corresponding identities imply a fractonic continuity equation, encoding the restricted mobility of isolated charges and the motion of dipoles. This justifies referring to the covariant higher-rank Chern-Simons model as a fractonic theory.

The purpose of the present paper is to determine the edge structures induced by a planar boundary in this covariant higher-rank Chern-Simons theory. The boundary observables provide characteristic low-energy information about the
higher-rank gauge structure: the current algebras fix the universal central/contact structure of the equal-time edge-current algebra, without by themselves fixing the full correlators, while the boundary Hamiltonians determine the propagation velocities, and the different derivative orders of the central terms distinguish the ordinary and fractonic current sectors. In a microscopic realisation, these features can appear in edge-current response and spectral functions and may therefore provide experimental discriminators of the higher-rank boundary dynamics. The induced bosonic actions also establish the connection with $(1+1)$-dimensional conformal field theory and provide the boundary data needed for entanglement calculations. Since the bulk theory is not topological, the relation between edge and bulk observables is itself part of the physics. We also address the compatibility of the mixed bulk-boundary commutator with physical microcausality. In a gauge theory, this condition must be tested on local gauge-invariant observables. In the present higher-rank theory these observables are generated by the field strength that generalises the Maxwell tensor to the symmetric rank-two gauge field and by its derivatives. We therefore analyse the corresponding bulk response and determine how the boundary current algebra acts on the local gauge-invariant observables.

More generally, boundary degrees of freedom arise naturally when local subsystems are defined in gauge theories and gravity \cite{Donnelly:2016auv}. In ordinary Chern-Simons theory, the boundary turns gauge invariance into a broken Ward identity and produces a Kac-Moody algebra \cite{Witten:1988hf, Elitzur:1989nr, Geiller:2017xad}. The corresponding lower-dimensional dynamics is described by chiral bosons and Tomonaga-Luttinger liquids \cite{Wen:1990se, Maggiore:2017vjf, Maggiore:2018bxr, Blasi:2011pf}. The higher-rank theory studied here has a single scalar gauge symmetry. In the presence of the boundary, the scalar gauge generator singles out one functional field-equation combination whose zeroth and first normal moments are directly relevant to the edge: the zeroth moment involves the tangential tensor $h_{ij}$, whereas the first moment also retains the normal-tangential field $w_i=h_{2i}$.

Our analysis follows the boundary quantum field theory approach initiated by Symanzik \cite{Symanzik:1981wd} and developed for boundary reductions of topological field theories \cite{Amoretti:2014iza}. The most general local boundary term compatible with power counting is included in the action, and the variational principle determines the boundary conditions. The zeroth and first normal moments of this scalar functional relation then determine the two boundary currents and their algebras.

The main results are the following. The two integrated boundary identities yield a traceless rank-two current with a fractonic conservation law and an ordinary vector current. Their functional derivatives give, respectively, a higher-rank Kac-Moody algebra and a standard $U(1)$ Kac-Moody algebra. The mixed spatial-current bulk-boundary commutator is proportional to the normal distance and vanishes when the bulk insertion is brought to the boundary. The equal-time spatial field-strength commutators vanish without any restriction on the boundary parameters, thus satisfying the microcausality condition on the equal-time spacelike slice. On the decoupled branch $c^{ijk}=0$ of the boundary conditions, and within the same local non-zero-spatial-momentum sector, the full unequal-time mixed commutator is annihilated by all field-strength components and the corresponding retarded response vanishes at all times, thereby establishing microcausality in this mixed bulk--boundary response channel. On the same branch, the induced action is the direct sum of a chiral scalar $\chi$ and the higher-rank sector described by $\xi$ and the auxiliary field $\phi$; on this branch the intrinsic boundary symplectic form is block diagonal. After eliminating $\phi$, the field $\psi=\partial_1\xi$ has a Luttinger-liquid action on the positive-energy branch. The bulk boundary conditions fix the chiral $w_i$ dynamics and select the positive-energy chiral sector of the $h_{ij}$ theory. The positivity conditions imply that the two matched chiral modes co-propagate, with both velocities having sign opposite to $\kappa$. Finally, the geometric decomposition on a non-planar boundary shows how extrinsic curvature mixes tangential and normal projections. Because this embedding data enters already in the projected gauge structure, it can provide a geometric distinction from the standard intrinsic Luttinger description that is hidden in the planar limit.

The paper is organised as follows. In Section~2 we introduce the covariant higher-rank Chern-Simons action and its planar-boundary formulation. In Section~3 we derive the equations of motion and boundary conditions. Section~4
formulates the scalar Ward combination of the functional field equations and derives its zeroth and first normal moments. In Section~5 we derive the higher-rank, ordinary and mixed equal-time algebras. Section~6 constructs the induced boundary theory and discusses its Luttinger, chiral and physical interpretations. Section~7 matches both boundary sectors to the bulk boundary conditions on the decoupled branch $c^{ijk}=0$. Section~8 analyses microcausality through the response of the local gauge-invariant field strength. Section~9 discusses non-planar boundaries, and Section~10 contains our conclusions.

\medskip
\noindent{\bf Notations and conventions.}
We work in $(2+1)$-dimensional spacetime with coordinates $x^\mu=(x^0,x^1,x^2)$. Greek indices $\mu,\nu,\rho,\ldots=0,1,2$ label spacetime components. After introducing the boundary at $x^2=0$, we denote the boundary coordinates by $X^i=(x^0,x^1)$ and use Latin indices $i,j,k,\ldots=0,1$ for components parallel to the boundary. The metric follows the mostly-plus convention $\eta_{\mu\nu}=\mathrm{diag}(-1,1,1)$, and the Levi-Civita tensor is normalised as $\epsilon^{012}=1=-\epsilon_{012}$. We set $\epsilon^{ij}\equiv\epsilon^{ij2}$. For a non-planar boundary, the normal is chosen spacelike and normalised by $n_\mu n^\mu=+1$. 

%**********************************************
\section{The Model}
%**********************************************

To construct the action in three spacetime dimensions, we look for the most general local functional for a symmetric gauge field $h_{\mu\nu}$ that is linear in derivatives and invariant under
\begin{equation}
    \delta h_{\mu\nu} = \partial_\mu\partial_\nu\lambda\,,\label{fracton_gauge}
\end{equation}
which is a particular case of linear diffeomorphisms 
\be
\delta_{diff}h_{\mu\nu}=\partial_\mu\xi_\nu+\partial_\nu\xi_\mu\qquad;\qquad
\xi_\mu=\frac{1}{2}\partial_\mu\lambda\ ,
\label{lindiff}\ee
known as longitudinal diffeomorphisms \cite{Dalmazi:2020qqn}. 
The unique covariant action satisfying these requirements is the higher-rank Chern-Simons form~\cite{Bertolini:2024yur}
\begin{equation}
    S_{\rm inv} = \kappa\int d^3x\,\epsilon^{\mu\nu\rho}h_{\mu\lambda}\partial_\nu h_\rho{}^\lambda\,,\label{S_inv}
\end{equation}
where $\kappa$ is a coupling constant which we keep explicit, since it will play an important role in what follows.
From \eqref{S_inv} one reads off the mass dimension $[h_{\mu\nu}]=1$.
The action \eqref{S_inv} is traceless \cite{Bertolini:2024yur}, since it depends only on the traceless component of the gauge field,
\begin{equation}
    \bar{h}_{\mu\nu} \equiv h_{\mu\nu} - \tfrac{1}{3}\eta_{\mu\nu}h\,,\label{h_bar}
\end{equation}
where $h\equiv\eta^{\mu\nu}h_{\mu\nu}$, so that $\eta^{\mu\nu}\bar{h}_{\mu\nu}=0$. 
It is easy to check that the invariant action \eqref{S_inv} can be written as
\begin{equation}
    S_{\rm inv} = \kappa\int d^3x\,\epsilon^{\mu\nu\rho}\bar{h}_{\mu\lambda}\partial_\nu\bar{h}_\rho{}^\lambda\,.
\end{equation}
We study this theory in the half-space $x^2\geq 0$, which can be implemented by inserting the Heaviside step function into the action:
\begin{equation}
    S_{\rm bulk} = \kappa\int d^3x\,\theta(x^2)\,\epsilon^{\mu\nu\rho}h_{\mu\lambda}\partial_\nu h_\rho{}^\lambda\,.\label{S_bulk}
\end{equation}
The six components of the symmetric tensor $h_{\mu\nu}$ can be organised as  \begin{equation}
    h_{22}\equiv\sigma\,,\qquad h_{2i}\equiv w_i\,,\qquad h_{ij}\,,
\end{equation}
with $i,j=0,1$. In terms of these fields,
\begin{equation}
    S_{\rm bulk} = \kappa\int d^3x\,\theta(x^2)\,\epsilon^{ij}\Bigl[2w_i\partial_j\sigma + 2h_i{}^k\partial_j w_k - \bigl(w_i\partial_2 w_j + h_{ik}\partial_2 h_j{}^k\bigr)\Bigr]\,,\label{S_bulk_decomp}
\end{equation}
where $\epsilon^{ij}\equiv\epsilon^{ij2}$.

Since the gauge parameter $\lambda$ is a scalar, the natural gauge-fixing condition is
the axial one
\begin{equation}
    \sigma = 0\,,
\end{equation}
enforced through a Lagrange multiplier $b$:
\begin{equation}
    S_{\rm gf} = \int d^3x\,\theta(x^2)\,b\,\sigma\,.
\end{equation}
Moreover, external sources $J^{ij}$ and $J^i$ are coupled to $h_{ij}$ and $w_i$, respectively:
\begin{equation}
    S_J = \int d^3x\,\theta(x^2)\bigl[J^{ij}h_{ij} + J^i w_i\bigr]\,.
\end{equation}
In Symanzik's approach \cite{Symanzik:1981wd}, boundary conditions are not imposed \emph{a priori} but are derived by adding to the bulk action the most general local boundary term compatible with power counting:
\begin{equation}
    S_{\rm bd} = \int d^3x\,\delta(x^2)\Bigl[c^{ij}w_i w_j + c^{ijk}w_i h_{jk} + c^{ijkl}h_{ij}h_{kl}\Bigr]\,,\label{S_bd}
\end{equation}
where the constant dimensionless coefficients satisfy
\begin{equation}
    c^{ij}=c^{ji}\,,\qquad c^{ijk}=c^{ikj}\,,\qquad c^{ijkl}=c^{jikl}=c^{ijlk}=c^{klij}\,.
\end{equation}
Hence, the total action of the theory is
\begin{equation}
    S_{\rm tot} = S_{\rm bulk} + S_{\rm gf} + S_J + S_{\rm bd}\,.\label{S_tot}
\end{equation}

%**********************************************
\section{Equations of motion and boundary conditions}
%**********************************************

The equations of motion derived from \eqref{S_tot} read
\begin{align}
\frac{\delta S_{\rm tot}}{\delta\sigma} &= \theta(x^2)\bigl(2\kappa\,\epsilon^{ij}\partial_i w_j + b\bigr) = 0\,,\\[6pt]
\frac{\delta S_{\rm tot}}{\delta w_i} &= \theta(x^2)\Bigl[2\kappa\bigl(\epsilon^{ij}(\partial_j\sigma - \partial_2 w_j) + \epsilon^{jk}\partial_j h_k{}^i\bigr) + J^i\Bigr]\notag\\
&\quad + \delta(x^2)\Bigl[\kappa\,\epsilon^{ji}w_j + 2c^{ij}w_j + c^{ijk}h_{jk}\Bigr] = 0\,,\label{EoM_w}\\[6pt]
\frac{\delta S_{\rm tot}}{\delta h_{ij}} &= \theta(x^2)\Bigl[\kappa\,\epsilon^{ik}(\partial_k w^j - \partial_2 h_k{}^j) + \kappa\,\epsilon^{jk}(\partial_k w^i - \partial_2 h_k{}^i) + J^{ij}\Bigr]\notag\\
&\quad + \delta(x^2)\Bigl[-\tfrac{\kappa}{2}(\epsilon^{ik}h_k{}^j + \epsilon^{jk}h_k{}^i) + c^{kij}w_k + 2c^{ijkl}h_{kl}\Bigr] = 0\,.\label{EoM_hij}
\end{align}
The boundary conditions (BC) are obtained by integrating each equation of motion across the boundary, $\lim_{\epsilon\to 0}\int_0^\epsilon dx^2\,(\mathrm{EoM}) = 0$, and we get the following system of coupled BC
\begin{eqnarray}
        \Bigl[\kappa\,\epsilon^{ji}w_j + 2c^{ij}w_j + c^{ijk}h_{jk}\Bigr]_{x^2=0} &=& 0\label{BC_w}\\
        \Bigl[-\tfrac{\kappa}{2}(\epsilon^{ik}h_k{}^j + \epsilon^{jk}h_k{}^i) + c^{kij}w_k + 2c^{ijkl}h_{kl}\Bigr]_{x^2=0} &=& 0\,.\label{BC_h}
\end{eqnarray}

%**********************************************
\section{Scalar functional Ward relation and normal moments}
%**********************************************

It is useful to organise the boundary analysis directly at the level of the functional field equations, while keeping explicit that the bulk gauge symmetry is a single scalar one. Denote by $E^{\mu\nu}$ the functional derivative of $S_{\rm tot}$ with respect to the symmetric gauge field $h_{\mu\nu}$. In terms of the variables used in the half-space decomposition,
\begin{equation}
 E^{ij}=\frac{\delta S_{\rm tot}}{\delta h_{ij}}\,,\qquad
 2E^{i2}=\frac{\delta S_{\rm tot}}{\delta w_i}\,,\qquad
 E^{22}=\frac{\delta S_{\rm tot}}{\delta\sigma}\,.
 \label{E_components}
\end{equation}
The field equations are $E^{\mu\nu}=0$. The formal adjoint of the scalar
gauge generator $\delta h_{\mu\nu}=\partial_\mu\partial_\nu\lambda$ is the
double-divergence operator $\partial_\mu\partial_\nu$. Applying this operator
to the functional field equations gives the scalar Ward combination
\begin{equation}
 \mathcal W[E]\equiv\partial_\mu\partial_\nu E^{\mu\nu}
 =\partial_i\partial_jE^{ij}
 +2\partial_i\partial_2E^{i2}
 +\partial_2^2E^{22}=0\,.
 \label{functional_WI_split}
\end{equation}
Equation~\eqref{functional_WI_split} is therefore a differential consequence of the functional field equations, selected by the same differential operator that generates the scalar gauge transformation; it is not an additional off-shell Noether identity of $S_{\rm tot}$, which contains the gauge-fixing, source and boundary terms. At the quantum level the equations are understood in the standard Schwinger--Dyson sense, so that source differentiation of the resulting integrated relations generates the correlator identities used below. The broken boundary identities appear after the explicit functional derivatives and boundary conditions are inserted into the normal projections. The Heaviside function defining the half-space is already contained in $S_{\rm tot}$ and therefore in the functional derivatives in \eqref{E_components}.

We first take the zeroth normal moment. Integrals without explicit limits in this functional step extend over the full normal line; after the functional derivatives are inserted, their support reduces them to the half-space and to possible terms at $x^2=0$. With the usual decay at normal infinity, the normal total derivatives in \eqref{functional_WI_split} integrate to zero, so that
\begin{equation}
 \int_{-\infty}^{+\infty}dx^2\,\partial_i\partial_jE^{ij}=0\,.
 \label{functional_WI_zeroth}
\end{equation}
Inserting Eq.~\eqref{EoM_hij}, using the boundary condition \eqref{BC_h}, and
integrating the normal derivative in the bulk term gives
\begin{equation}
 \int_0^\infty dx^2\,\partial_i\partial_jJ^{ij}
 =-2\kappa\,\epsilon^{ik}\partial_i\partial_jh_k{}^j\big|_{x^2=0}\,.
 \label{WI}
\end{equation}
This is the broken boundary identity associated with the tangential tensor sector.

The first normal moment retains the normal-tangential equation. Multiplying Eq.~\eqref{functional_WI_split} by $x^2$ before the normal integration gives
\begin{equation}
 \int dx^2\,x^2\partial_i\partial_jE^{ij}
 -2\partial_i\int dx^2\,E^{i2}=0\,.
 \label{functional_WI_first}
\end{equation}
Indeed,
\begin{equation}
 2\int dx^2\,x^2\partial_i\partial_2E^{i2}
 =-2\partial_i\int dx^2\,E^{i2}\,,
 \label{functional_mixed_moment}
\end{equation}
while $\int dx^2\,x^2\partial_2^2E^{22}=0$ provided the fields and sources decay sufficiently fast for the weighted integrals to converge and for all endpoint terms generated by integration by parts to vanish. Since $2E^{i2}=\delta S_{\rm tot}/\delta w_i$, the second term in \eqref{functional_WI_first} is precisely the integrated tangential divergence of the $w_i$ field equation. Substituting Eqs.~\eqref{EoM_hij} and \eqref{EoM_w}, and using the boundary conditions, one obtains
\begin{equation}
 \int_0^\infty dx^2\left[
 x^2\partial_i\partial_jJ^{ij}-\partial_iJ^i\right]
 =2\kappa\,\epsilon^{ij}\partial_iw_j\big|_{x^2=0}\,.
 \label{WI_w}
\end{equation}
Thus the normal total derivative that removes the $w_i$ contribution from the zeroth moment does not remove it from the first one. Equation~\eqref{WI_w} is exactly the bulk-plus-boundary relation obtained by combining the first normal
moment of the $h_{ij}$ equation with the integrated tangential divergence of the $w_i$ equation.

As a direct component check, under the same endpoint conditions, the double tangential divergence of the bulk part of Eq.~\eqref{EoM_hij} gives
\begin{equation}
 \partial_i\partial_jJ^{ij}
 =2\kappa\,\partial_2\!\left(
 \epsilon^{ik}\partial_i\partial_jh_k{}^j\right)\,.
 \label{WI_local_check}
\end{equation}
Its first normal moment yields
\begin{equation}
 \int_0^\infty dx^2\,x^2\partial_i\partial_jJ^{ij}
 =-2\kappa\int_0^\infty dx^2\,
 \epsilon^{ik}\partial_i\partial_jh_k{}^j\,.
 \label{WI_first_check}
\end{equation}
The integrated divergence of the bulk part of Eq.~\eqref{EoM_w} gives
\begin{equation}
 \int_0^\infty dx^2\,\partial_iJ^i
 =-2\kappa\,\epsilon^{ij}\partial_iw_j\big|_{x^2=0}
 -2\kappa\int_0^\infty dx^2\,
 \epsilon^{ik}\partial_i\partial_jh_k{}^j\,.
 \label{WI_w_check}
\end{equation}
Eliminating the common bulk integral between \eqref{WI_first_check} and \eqref{WI_w_check} reproduces Eq.~\eqref{WI_w}. This direct derivation also makes explicit the finite bulk contribution entering the second integrated identity.

The two boundary-relevant moments have a natural relation with the axial gauge. Preservation of $\sigma=h_{22}=0$ requires
\begin{equation}
 \partial_2^2\lambda=0\,.
 \label{residual_condition}
\end{equation}
Accordingly, along the normal direction a gauge-preserving parameter has the
affine form
\begin{equation}
 \lambda(X,x^2)=\lambda_0(X)+x^2\lambda_1(X)\,,
 \label{residual_lambda}
\end{equation}
where
\begin{equation}
 \lambda_0(X)\equiv\lambda|_{x^2=0}\,,\qquad
 \lambda_1(X)\equiv\partial_2\lambda|_{x^2=0}\,.
 \label{boundary_lambda_data}
\end{equation}
Thus $\lambda_0$ and $\lambda_1$ are respectively the boundary value and first normal derivative of the same scalar gauge parameter; they do not represent two independent bulk gauge symmetries. Equation~\eqref{residual_lambda}
characterises the local solutions of the gauge-preserving equation in the normal direction. We do not assume that every such affine solution defines an admissible global gauge transformation: a generic $\lambda_1(X)$ can spoil the required falloff as $x^2\to\infty$. The derivation of \eqref{WI} and \eqref{WI_w} uses only the scalar functional relation \eqref{functional_WI_split}, its first two normal moments, and the stated convergence conditions on fields and sources.

For completeness, higher normal moments do not produce further independent boundary breakings, assuming sufficiently rapid weighted falloff so that the corresponding moments converge and all endpoint terms generated by the
integrations by parts vanish. For $n\geq1$,
\begin{equation}
 2\int dx^2\,(x^2)^n\partial_i\partial_2E^{i2}
 =-2n\,\partial_i\int dx^2\,(x^2)^{n-1}E^{i2}\,,
 \label{higher_mixed_moments}
\end{equation}
and for $n\geq2$,
\begin{equation}
 \int dx^2\,(x^2)^n\partial_2^2E^{22}
 =n(n-1)\int dx^2\,(x^2)^{n-2}E^{22}\,.
 \label{higher_sigma_moments}
\end{equation}
For $n\geq2$ these relations involve bulk moments. After the integration by parts in \eqref{higher_mixed_moments}, the factor $(x^2)^{n-1}$ multiplying $E^{i2}$ eliminates any $\delta(x^2)$ contribution. Moreover $E^{22}$ has no
boundary-localised term because the boundary action \eqref{S_bd} is independent of $\sigma$. Hence moments above first order relate bulk integrals and do not supply additional independent breakings localised at the boundary.

At vanishing external sources, Eqs.~\eqref{WI} and \eqref{WI_w} give
\begin{equation}
 \epsilon^{ik}\partial_i\partial_jh_k{}^j\big|_{x^2=0}=0\,,\qquad
 \epsilon^{ij}\partial_iw_j\big|_{x^2=0}=0\,.
 \label{on_shell}
\end{equation}
These are conservation laws for the boundary currents
\begin{equation}
 \mathcal J^{ij}\equiv
 \epsilon^{ik}h_k{}^j+\epsilon^{jk}h_k{}^i\,,\qquad
 \mathcal K^i\equiv\epsilon^{ij}w_j\,,
 \label{boundary_currents}
\end{equation}
namely
\begin{equation}
 \partial_i\partial_j\mathcal J^{ij}=0\,,\qquad
 \partial_i\mathcal K^i=0\,.
 \label{current_conservation}
\end{equation}
The first can be written in the standard scalar-charge continuity form~\cite{Pretko:2016lgv} by
defining
\begin{equation}
 \rho_{\partial}\equiv
 \partial_0\mathcal J^{00}+2\partial_1\mathcal J^{01}\,,\qquad
 \partial_0\rho_{\partial}+\partial_1^2\mathcal J^{11}=0\,.
 \label{boundary_fracton_continuity}
\end{equation}
Under the standard spatial falloff, Eq.~\eqref{boundary_fracton_continuity} implies conservation of the boundary charge $Q_{\partial}=\int dx^1\,\rho_{\partial}$ and dipole moment $D_{\partial}=\int dx^1\,x^1\rho_{\partial}$. The second relation in \eqref{current_conservation} is an ordinary current conservation law. The rank-two current is traceless, $\eta_{ij}\mathcal J^{ij}=0$, and depends only on the traceless component of $h_{\mu\nu}$. The planar boundary therefore carries both the fractonic current of the tangential tensor sector and the ordinary current associated with the normal-tangential field.

%**********************************************

\section{Boundary algebras}
%**********************************************

\subsection{Higher-rank algebra}

The boundary algebra induced by the zeroth-moment boundary identity is obtained by differentiating \eqref{WI} with respect to the external source $J^{mn}$ and then setting $J=0$:
\begin{equation}
\left.\frac{\delta}{\delta J^{mn}(x')}\int_0^\infty dx^2\,\partial_i\partial_j J^{ij}\right|_{J=0}
= -2\kappa\,\epsilon^{ik}\partial_i\partial_j\left.\frac{\delta h_k{}^j(x)}{\delta J^{mn}(x')}\right|_{x^2=0,\,J=0}\,.\label{der_WI}
\end{equation}
The left-hand side gives
\begin{align}
\int_0^\infty dx^2\,\partial_i\partial_j\,\frac{\delta J^{ij}(x)}{\delta J^{mn}(x')}\bigg|_{J=0}
&= \int_0^\infty dx^2\,\partial_i\partial_j\left(\delta^3(x-x')\,\tfrac{1}{2}(\delta_i^m\delta_j^n+\delta_i^n\delta_j^m)\right) \notag\\
&= \partial_m\partial_n\,\delta^2(X-X')\,.\label{lhs}
\end{align}
Defining the time-ordered product as
\begin{equation}
T\!\left[h_{ij}(x)h_{mn}(x')\right]
\equiv
\theta(x^0-{x'^0})\,h_{ij}(x)h_{mn}(x')+\theta({x'^0}-x^0)\,h_{mn}(x')h_{ij}(x)\ ,
\end{equation}
and using $h_{mn}(x')=\delta Z_c/\delta J^{mn}(x')$, the right-hand side of \eqref{der_WI} involves
\begin{equation}
-2\kappa\,\epsilon^{ik}\partial_i\partial^j \frac{\delta^2 Z_c}{\delta J^{mn}(x')\,\delta J^{kj}(x)}\bigg|_{\substack{x^2=0\\J=0}}
=
-2i\kappa\,\epsilon^{ik}\partial_i\partial^j\bigl\langle T\,h_{kj}(x)h_{mn}(x')\bigr\rangle\big|_{x^2=0}\,.\label{dZc}
\end{equation}
The evaluation of this expression proceeds via the Schwinger-Dyson identity. For any two operators $A(x)$ and $B(x')$, differentiating the $\theta$-functions in the definition of the time-ordered product gives
\begin{equation}
\partial_\mu^x\langle T[A(x)\,B(x')]\rangle
=
\langle T[\partial_\mu A(x)\,B(x')]\rangle
+\delta_\mu^{0}\,\delta(x^0-{x'^0})\,[A(x),\,B(x')]\big|_{x^0={x'^0}}\,,
\label{SD}
\end{equation}
where only time derivatives produce contact terms. Applying \eqref{SD} repeatedly to \eqref{dZc} separates the result into a bulk term and two contact terms:
\begin{align}
-2i\kappa\epsilon^{ik}\partial_i\partial^j\langle T\,h_{kj}\,h'_{mn}\rangle\big|_{x^2=0}
&=
-2i\kappa\underbrace{\langle T\bigl[\epsilon^{ik}\partial_i\partial^j h_{kj}\bigr]h'_{mn}\rangle\big|_{x^2=0}}_{\displaystyle=\;0\;\text{by \eqref{on_shell}}}
\notag\\
&\quad -2i\kappa\,\delta(x^0-{x'^0})\bigl[\partial^j h_{j1}+\epsilon^{ik}\partial_i h^0{}_{k},h_{mn}'\bigr]_{x^2=0}
\notag\\
&\quad-2i\kappa\,\partial_0\delta(x^0-{x'^0})\bigl[h^0_{\ 1},h_{mn}'\bigr]_{x^2=0}
\notag\\[4pt]
&=-2i\kappa\;\delta(x^0-{x'^0})\,\bigl[\partial_1(h_{00}+h_{11})-2\partial_0 h_{01},\,h_{mn}'\bigr]_{x^2=0}
\notag\\[4pt]
&\quad+2i\kappa\;\partial_0\delta(x^0-{x'^0})\,\bigl[h_{01},\,h_{mn}'\bigr]_{x^2=0}\,.
\label{contact_terms}
\end{align}
Combining \eqref{lhs} with \eqref{contact_terms} and integrating over $x^0$, the $\partial_0\delta$ term gives $[\partial_0h_{01},h_{mn}']\delta(x^0-x'^0)$. One obtains
\begin{equation}
\int dx^0\,\partial_m\partial_n\bigl(\delta(x^0-x'^0)\delta(x^1-x'^1)\bigr)
= -2i\kappa\,\bigl[(\partial_1(h_{00}+h_{11}) - \partial_0 h_{01})(X),\,
h_{mn}(X')\bigr]_{x^0=x'^0}\,.\label{pre_algebra}
\end{equation}
For the purely spatial insertion $m=n=1$, for which the left-hand side contains no time derivative of the contact delta function, one obtains the equal-time boundary algebra
\begin{equation}
    \bigl[(\partial_1(h_{00}+h_{11}) - \partial_0 h_{01})(X),\,h_{11}(X')\bigr]_{x^0={x'}^0}
    = \frac{i}{2\kappa}\,\partial_1^2\,\delta(x^1-{x'}^1)\,.\label{comm_rule}
\end{equation}
This algebra is the higher-rank analogue of the boundary Kac-Moody algebra arising in ordinary Chern-Simons theory \cite{Witten:1988hf, Elitzur:1989nr}. A closely related generalised $U(1)$ Kac-Moody algebra was found in the boundary analysis of the (3+1)-dimensional covariant fracton gauge theory of Ref.~\cite{Bertolini:2023sqa}. The same boundary mechanism also produces standard Kac-Moody structures in linearized gravity with boundary \cite{Bertolini:2023wie}. In the present case this mechanism appears in a purely (2+1)-dimensional higher-rank Chern-Simons setting: the breaking of the Ward identity by the boundary source terms induces a non-trivial equal-time algebra for the boundary fields.

\subsection{Ordinary and mixed algebras}

Differentiating the first-normal-moment identity \eqref{WI_w} with respect to $J^m(x')$ and setting the sources to zero gives
\begin{equation}
 -\partial_m\delta^2(X-X')
 =2i\kappa\,\epsilon^{ij}\partial_i
 \left\langle T\,w_j(X)w_m(x')\right\rangle\,.
 \label{der_WI_w}
\end{equation}
Choosing the purely spatial insertion $m=1$, taking it to the boundary and using $\epsilon^{ij}\partial_iw_j=\partial_0w_1-\partial_1w_0=0$, the time derivative produces the equal-time contact term. Integration over $x^0$ gives
\begin{equation}
 \bigl[w_1(X),w_1(X')\bigr]_{x^0=x'^0}
 =\frac{i}{2\kappa}\,\partial_1\delta(x^1-x'^1)\,.
 \label{w_algebra}
\end{equation}
Since $\mathcal K^0=w_1$, this is the standard $U(1)$ Kac-Moody algebra of the ordinarily conserved boundary current.

The mixed relation follows by differentiating \eqref{WI_w} with respect to $J^{mn}(x')$, where $x'=(X',x^{\prime 2})$:
\begin{equation}
 x^{\prime 2}\partial_m\partial_n\delta^2(X-X')
 =2i\kappa\,\epsilon^{ij}\partial_i
 \left\langle T\,w_j(X)h_{mn}(x')\right\rangle\,.
 \label{mixed_derivative}
\end{equation}
For the purely spatial insertion $m=n=1$, the equal-time relation is
\begin{equation}
 \bigl[w_1(X),h_{11}(x')\bigr]_{x^0=x'^0}
 =-\frac{i\,x^{\prime 2}}{2\kappa}\,\partial_1^2
 \delta(x^1-x'^1)\,.
 \label{mixed_bulk_boundary}
\end{equation}
This is a non-zero bulk-boundary commutator at finite normal distance. Its intrinsic boundary limit is
\begin{equation}
 \bigl[w_1(X),h_{11}(X')\bigr]_{x^0=x'^0}=0\,.
 \label{mixed_boundary_zero}
\end{equation}
Thus Eq.~\eqref{mixed_bulk_boundary} relates the boundary current $w_1$ to the gauge-fixed bulk component $h_{11}$: the commutator is non-zero at finite normal distance and vanishes in the intrinsic boundary limit. Its projection onto the local gauge-invariant field strength, which determines the corresponding physical response, is evaluated in Section~\ref{section_micro} and vanishes, so microcausality is satisfied.

%**********************************************
\section{Induced boundary theory}\label{section_2D}
%**********************************************

The two on-shell constraints in \eqref{on_shell} provide two local boundary parametrisations. The ordinary conservation law gives
\begin{equation}
 \epsilon^{ij}\partial_iw_j=0
 \quad\Longrightarrow\quad
 w_i=\partial_i\chi\,,
 \label{w_param}
\end{equation}
where $\chi$ is a dimensionless scalar, up to global zero modes. The higher-rank constraint parametrises $h_{ij}$ in terms of $\xi$ and $\phi$, as shown below. The mixed intrinsic spatial current commutator in Eq.~\eqref{mixed_boundary_zero} vanishes. The identities and spatial current algebras derived above do not require any restriction on $c^{ijk}$. For the explicit boundary dynamics and bulk--boundary matching we specialise to the decoupled branch $c^{ijk}=0$, on which the boundary conditions separate and the induced action is
\begin{equation}
 S_{\partial}=S_w[\chi]+S_h[\xi,\phi] \,.
 \label{complete_boundary_action}
\end{equation}
The absence of mixed kinetic terms in Eq.~\eqref{complete_boundary_action} then makes the intrinsic boundary symplectic form block diagonal on this branch.

\subsection{The \texorpdfstring{$w_i$}{wi} sector}

In terms of $\chi$, Eq.~\eqref{w_algebra} becomes
\begin{equation}
 \bigl[\partial_1\chi(X),\partial_1\chi(X')\bigr]_{x^0=x'^0}
 =\frac{i}{2\kappa}\,\partial_1\delta(x^1-x'^1)\,.
 \label{chi_algebra}
\end{equation}
The primary constraint associated with the symplectic term is
\begin{equation}
 \Omega_\chi=\pi_\chi-\kappa\partial_1\chi=0\,,\qquad
 \{\Omega_\chi(X),\Omega_\chi(X')\}
 =-2\kappa\,\partial_1\delta(x^1-x'^1)\,.
 \label{chi_constraint}
\end{equation}
Its Dirac bracket reproduces \eqref{chi_algebra}. The local quadratic first-order action with this symplectic structure and invariant under constant shifts of $\chi$ is
\begin{equation}
 S_w[\chi]=\int d^2X\left[
 \kappa\,\partial_1\chi\,\partial_0\chi
 +a_w(\partial_1\chi)^2\right] ,
 \label{S_w}
\end{equation}
where $a_w$ is dimensionless. Its equation of motion is
\begin{equation}
 \partial_1\left(\kappa\,\partial_0\chi
 +a_w\partial_1\chi\right)=0\,.
 \label{EoM_chi}
\end{equation}
After removing the spatial zero mode, this is a chiral equation with velocity
$v_w=a_w/\kappa$. The Hamiltonian density is
\begin{equation}
 \mathcal H_w=-a_w(\partial_1\chi)^2\,,
 \label{H_w}
\end{equation}
and its positive-energy branch is $a_w<0$.

\subsection{The higher-rank \texorpdfstring{$h_{ij}$}{hij} sector}

At vanishing external sources, the zeroth-moment identity \eqref{WI} yields the boundary constraint
\begin{equation}
    \epsilon^{ik}\partial_i\partial^j h_{kj}\big|_{x^2=0} = 0\,,\label{bdconstraint}
\end{equation}
which can be solved as follows. Rewriting it as
\begin{equation}
    \partial^j\bigl(\epsilon^{ik}\partial_i h_{kj}\big|_{x^2=0}\bigr)=0
    \quad\Longrightarrow\quad
    \epsilon^{ik}\partial_i h_{kj}\big|_{x^2=0} = \epsilon_{ij}\,\partial^i\phi\,,\label{step_1}
\end{equation}
where $\phi$ is a generic $(1+1)$-dimensional scalar function. This step must be understood locally on the boundary: in two spacetime dimensions a divergence-free vector can be written locally as the dual of the gradient of a scalar, up to global contributions or zero modes not captured by the above local representation. From \eqref{step_1} it follows that
\begin{equation}
    \epsilon^{ik}\partial_i\bigl(h_{kj}\big|_{x^2=0} - \eta_{jk}\phi\bigr) = 0\,,\label{step_2}
\end{equation}
so that, locally, for each $j$ there exists a vector function $v_j$ such that
\begin{equation}
    h_{kj}\big|_{x^2=0} - \eta_{jk}\phi = \partial_k v_j\,.
\end{equation}
Since the left-hand side is symmetric in $k,j$, this implies $\partial_k v_j = \partial_j v_k$, and therefore, locally, there exists a scalar function $\xi$ such that $v_k = \partial_k\xi$. We conclude that
\begin{equation}
    h_{ij}\big|_{x^2=0} = \partial_i\partial_j\xi + \eta_{ij}\phi\,,\label{defhx2=0}
\end{equation}
$i.e.$\ the boundary field $h_{ij}$ admits a local parametrisation in terms of two scalar fields $\xi$ and $\phi$, with mass dimensions $[\xi]=-1$ and $[\phi]=+1$. This corresponds to the fact that, since $h_{ij}$ satisfies the differential constraint \eqref{bdconstraint}, its three components can be parametrised in terms of two scalar functions. The physical propagating degree of freedom is identified below. The scalar parametrisation \eqref{defhx2=0} is not unique. It is left
unchanged by the intrinsic redundancy
\begin{equation}
    \delta\xi = a + b_i X^i - \tfrac{c}{2}X^iX_i\,,\qquad \delta\phi = c\,,\label{trasf_xi_phi}
\end{equation}
where $a$, $b^i$ and $c$ are constant parameters with mass dimensions $-1$, $0$ and $+1$, respectively. Indeed, using $\partial_i\partial_j(X^kX_k)=2\eta_{ij}$, one has $\delta h_{ij}=\partial_i\partial_j\delta\xi+\eta_{ij}\delta\phi=0$. This is a redundancy of the scalar parametrisation of a fixed $h_{ij}$ and should not be confused with the scalar bulk gauge symmetry or with its boundary data $\lambda_0$ and $\lambda_1$. Expressing the algebra \eqref{comm_rule} in terms of the boundary scalars yields the equal-time commutation relation
\begin{equation}
    \Bigl[2\kappa\,\partial_1\xi(X),\,\partial_1^{\prime\,2}\xi(X')+\phi(X')\Bigr] = i\,\delta(x^1-x'^1)\,,\label{ccr_fields}
\end{equation}
which, once compared with the canonical commutation relations
\begin{equation}
    [q(X),p(X')] = i\,\delta(x^1-x'^1)\,,\label{ccr}
\end{equation}
identifies the canonical pair
\begin{equation}
    q \equiv 2\kappa\,\partial_1\xi\,,\qquad p \equiv \partial_1^2\xi + \phi\,,\label{canonical_pair}
\end{equation}
with $q$ playing the role of canonical coordinate and $p$ that of conjugate momentum. The dynamical part of the induced $(1+1)$-dimensional action is then
\begin{equation}
    S_{\rm kin} = \int d^2X\,p\,\dot q = 2\kappa\int d^2X\,(\partial_1^2\xi+\phi)\,\partial_0\partial_1\xi\,,\label{S_kin}
\end{equation}
to which one adds potential terms so that the total boundary action $S_h$ is
\begin{itemize}
    \item built from the canonical coordinate $q$ and its conjugate momentum $p$ in \eqref{canonical_pair};
    \item compatible with power counting;
    \item compatible with the parametrisation redundancy \eqref{trasf_xi_phi}: the local equations are invariant, while the action can change only by tangential total derivatives and field-independent terms;
    \item compatible with the canonical commutation relations \eqref{ccr}.
\end{itemize}
We identify actions that differ by tangential total derivatives or field-independent terms. With this convention, the most general local quadratic first-order phase-space action satisfying these requirements is
\begin{equation}
    S_h[p,q] = \int d^2X\left[p\,\dot q + \frac{a_1-a_2}{4\kappa^2}
    (\partial_1 q)^2 + a_2\,p^2\right]\,,\label{S_2D_pq}
\end{equation}
with $a_1$, $a_2$ dimensionless constants. The form \eqref{S_2D_pq} is a first-order action on phase space, in which $p$ and $q$ are independent variables: the term $p\,\dot{q}$ is the symplectic term of the Hamiltonian formulation, and $p$ is the canonical momentum conjugate to $q$, not a time derivative of $q$. Identifying
\begin{equation}
    S_h[q,p] = \int d^2X\,\bigl(p\,\dot q - \mathcal H(p,q)\bigr)\,,
\end{equation}
the Hamiltonian density reads
\begin{equation}
    \mathcal H(p,q)
    = -a_2\,p^2
    -\frac{a_1-a_2}{4\kappa^2}(\partial_1 q)^2\,.
    \label{H_density}
\end{equation}
Therefore, positive definiteness on the local non-zero-mode phase space is equivalent to
\begin{equation}
    a_2<0\,,\qquad a_1-a_2<0\,.
    \label{H_positivity}
\end{equation}
Writing \eqref{S_2D_pq} in terms of $\xi$ and $\phi$ via \eqref{canonical_pair}, we get
\begin{equation}
    S_h = \int d^2X\,\Bigl[2\kappa(\partial_1^2\xi+\phi)\,\partial_0\partial_1\xi
    + a_1(\partial_1^2\xi)^2 + a_2\,\phi\bigl(\phi+2\partial_1^2\xi\bigr)\Bigr]\,.\label{S_2D}
\end{equation}
This off-shell formulation contains higher derivatives and a field \(\xi\) with negative mass dimension, indicating that \(\xi\) and \(\phi\) are not the most appropriate dynamical variables for the effective description of the boundary. Indeed, from \eqref{S_2D} one sees that \(\phi\) appears without derivatives and is therefore an auxiliary field. Its algebraic equation of motion,
\begin{equation}
\frac{\delta S_h}{\delta\phi}=0\quad\Longrightarrow\quad
2\kappa\,\partial_0\partial_1\xi
+2a_2\left(\partial_1^2\xi+\phi\right)=0\,,
\label{EoM_phi}
\end{equation}
gives, for \(a_2\neq 0\),
\begin{equation}
\phi
=
-\frac{\kappa}{a_2}\,\partial_0\partial_1\xi
-\partial_1^2\xi\,.
\label{phi_on_shell}
\end{equation}
It is then more convenient to go back to the first-order formulation and integrate out $p$ directly. Its equation of motion is
\begin{equation}
\frac{\delta S_h}{\delta p}=0\quad\Longrightarrow\quad
\dot q + 2a_2\,p=0\quad\Longrightarrow\quad p=-\frac{\dot q}{2a_2}\,,
\end{equation}
again in the non-degenerate sector $a_2\neq 0$. The case $a_2=0$ is excluded, since $p$ would then act as a Lagrange multiplier enforcing $\dot q=0$ and thus trivialising the dynamics. Substituting $p=-\dot q/(2a_2)$ back into \eqref{S_2D_pq} gives
\begin{equation}
    S_h[q] = \int d^2X\left[-\frac{1}{4a_2}\,\dot q^{\,2} 
    + \frac{a_1-a_2}{4\kappa^2}(\partial_1 q)^2\right]\,,
\end{equation}
which is the desired second-order Lagrangian formulation: it is quadratic in derivatives and the field $q$ is dimensionless. The boundary case $a_1=a_2$ is excluded by \eqref{H_positivity}: the spatial gradient term vanishes, and although $\mathcal{H}=-a_2\,p^2\geq 0$ for $a_2<0$, the Hamiltonian density is only positive semidefinite, vanishing for any time-independent configuration. The Hamiltonian obtained from this second-order form selects the same positivity branch \eqref{H_positivity}. Introducing
\begin{equation}
    \psi \equiv \partial_1\xi\,,\qquad q = 2\kappa\,\psi\,,
\end{equation}
the induced action takes the second-order Lagrangian form
\begin{equation}
    S_h[\psi] = \int d^2X\left[-\frac{\kappa^2}{a_2}\,\dot\psi^2 + (a_1-a_2)(\partial_1\psi)^2\right]\,,\label{S_2D_psi}
\end{equation}
where $\phi$ has disappeared, as expected for an auxiliary field. The Hamiltonian associated with \eqref{S_2D_psi} is positive on the same branch \eqref{H_positivity}. The boundary dynamics is therefore described by a Gaussian second-order theory for the dimensionless field $\psi=\partial_1\xi$, which is the unique propagating degree of freedom in the $h_{ij}$ sector.

%------------------------------------------------------
\subsection{Connection with the Luttinger liquid}
%------------------------------------------------------

The action \eqref{S_2D_psi} coincides with the standard description of a non-chiral Luttinger liquid for the quasiparticle density mode \cite{Haldane:1981zza}
\begin{equation}
    \rho = \rho_+ + \rho_- \sim \partial_1\psi\,,\label{densities}
\end{equation}
where \(\rho_\pm(X)\) are the chiral densities. To see this, start from the chiral Luttinger liquid actions
\begin{equation}
    S_\pm = -k\int d^2X\,
    \partial_1\varphi_\pm
    \left(\pm\partial_0 + v\,\partial_1\right)\varphi_\pm\,,
    \label{S_chiral}
\end{equation}
where \(\varphi_+\) and \(\varphi_-\) propagate in opposite directions. Introducing
\begin{equation}
    \psi \equiv \frac{\varphi_-+\varphi_+}{\sqrt{2}}\,,
    \qquad
    \zeta \equiv \frac{\varphi_--\varphi_+}{\sqrt{2}}\,,
    \label{dual_fields}
\end{equation}
one obtains the equal-time algebra
\begin{equation}
[\zeta(X),\zeta(X')]
=
[\psi(X),\psi(X')]
=0\,,
\qquad
[\zeta(X),\partial_1'\psi(X')]
=
\frac{i}{2k}\,\delta(x^1-x'^1)\,.
\label{comm_dual}
\end{equation}
Thus the canonical pair is \(\{\zeta,\,2k\,\partial_1\psi\}\): in the harmonic-fluid terminology, \(\zeta\) is the phase field canonically conjugate to the density associated with \(2k\,\partial_1\psi\). In terms of $\psi$ and $\zeta$, integration by parts brings the total action $S=S_++S_-$ to the form
\begin{equation}
    S = -k\int d^2X\Bigl[v\bigl((\partial_1\zeta)^2+(\partial_1\psi)^2\bigr) - 2\,\partial_1\zeta\,\partial_0\psi\Bigr]\,.
\end{equation}
The equation of motion for $\zeta$ is $\partial_1(v\,\partial_1\zeta-\partial_0\psi)=0$, namely
\begin{equation}
    v\,\partial_1\zeta - \partial_0\psi = f(x^0)\,,
\end{equation}
where $f(x^0)$ is independent of $x^1$, $i.e.$\ it corresponds to the zero mode along $x^1$. Discarding this zero mode, or assuming boundary conditions that eliminate it, one obtains $v\,\partial_1\zeta=\partial_0\psi$. Substituting this relation back into the action yields
\begin{align}
S&=-k\int d^2X\left[v(\partial_1\psi)^2+v(\partial_1\zeta)^2-2\,\partial_1\zeta\,\partial_0\psi\right]\notag\\
&=-k\int d^2X\left[v(\partial_1\psi)^2+\frac{1}{v}(\partial_0\psi)^2-\frac{2}{v}(\partial_0\psi)^2\right]\notag\\
&=-k\int d^2X\left[-\frac{1}{v}(\partial_0\psi)^2+v(\partial_1\psi)^2\right]\,,
\end{align}
and therefore
\begin{equation}
    S = -k\int d^2X\left[-\frac{1}{v}(\partial_0\psi)^2 + v(\partial_1\psi)^2\right]\,,\label{S_Luttinger}
\end{equation}
which coincides with \eqref{S_2D_psi}, once we set
\begin{equation}
    v^2 = \frac{a_2(a_1-a_2)}{\kappa^2}\,,\qquad k^2 = \kappa^2\,\frac{a_1-a_2}{a_2}\,,\label{vk_id}
\end{equation}
which are both positive on the branch \eqref{H_positivity}. We note that the conditions $v^2>0$ and $k^2>0$ are each satisfied on both branches $\{a_2<0,\,a_1-a_2<0\}$ and $\{a_2>0,\,a_1-a_2>0\}$, since they depend only on the product $a_2(a_1-a_2)$ and the ratio $(a_1-a_2)/a_2$, both of which are positive on either branch. The physical branch \eqref{H_positivity} is therefore singled out by positive definiteness on the local non-zero-mode phase space alone. From the chiral densities $\rho_\pm\sim\partial_1\varphi_\pm$ and the definition \eqref{dual_fields}, one immediately recovers \eqref{densities}. The identification of $\psi$ with the total density mode follows the standard harmonic-fluid description of one-dimensional quantum liquids \cite{Haldane:1981zza}. In this description, $\psi$ is a displacement field: its spatial derivative measures the local fluctuation of the overall density, $i.e.$ the compressions and rarefactions of the quasiparticle fluid. Moreover, the algebra \eqref{comm_dual} shows that the variable canonically conjugate to the phase field \(\zeta\) is \(2k\,\partial_1\psi\). Thus \(\partial_1\psi\) describes the total density mode, while \(\psi\) is the
corresponding bosonic displacement field. In this sense, the action \eqref{S_2D_psi} describes a non-chiral system, with modes propagating in both directions.

%------------------------------------------------------
\subsection{Chiral sector}
%------------------------------------------------------

It is useful to distinguish explicitly two different ways of realising the condition $\phi=0$, since they highlight complementary aspects of the boundary dynamics and lead to different contacts with previous analyses of fractonic boundary modes.

The first realisation is obtained on-shell, by starting from the full theory \eqref{S_2D} and imposing $\phi=0$ only \emph{after} writing the equations of motion. From Eq.~\eqref{phi_on_shell} one obtains
\begin{equation}
    \phi = -\left(\frac{\kappa}{a_2}\,\partial_0 + \partial_1\right)\psi\,,
\end{equation}
so that $\phi=0$ is equivalent to the chirality equation
\begin{equation}\label{eq_chiral}
    \left(\partial_0 + \frac{a_2}{\kappa}\,\partial_1\right)\psi = 0\,,
\end{equation}
describing a chiral mode for $\psi$ with velocity $v_h = a_2/\kappa$. Compatibility with the equation of motion of $\xi$ further requires
\begin{equation}
    2\kappa\,\partial_0\partial_1^2\psi + a_1\,\partial_1^3\psi = 0\,,
\end{equation}
which, using \eqref{eq_chiral}, reduces to
\begin{equation}
    (a_1-2a_2)\,\partial_1^3\psi = 0\,.
\end{equation}
Up to the $x^1$-independent component (as in the case of the function $f(x^0)$ discussed above) or special boundary conditions, the non-trivial chiral subsector of the full theory therefore arises in a transparent way at
\begin{equation}
    a_1 = 2a_2\,.
\end{equation}

The second realisation is obtained at the level of the action, by imposing $\phi=0$ \emph{directly} in $S_h$\eqref{S_2D}. Using $\psi=\partial_1\xi$:
\begin{equation}
    S_h\big|_{\phi=0} = \int d^2X\Bigl[2\kappa\,\partial_1\psi\,\partial_0\psi + a_1(\partial_1\psi)^2\Bigr]\,.\label{S_phi0}
\end{equation}
The associated equation of motion,
\begin{equation}\label{eomPhi=0}
    \partial_1\!\left(2\kappa\,\partial_0\psi + a_1\,\partial_1\psi\right) = 0\,,
\end{equation}
becomes, in terms of the density $\rho$ introduced in \eqref{densities}, the continuity equation
\begin{equation}
    \partial_0\rho + \frac{a_1}{2\kappa}\,\partial_1\rho = 0\,,
\end{equation}
describing a chiral mode propagating with velocity $a_1/(2\kappa)$. The Hamiltonian density of \eqref{S_phi0} is
\begin{equation}
 \mathcal H_h=-a_1(\partial_1\psi)^2\,.
 \label{H_h_chiral}
\end{equation}
Its positivity requires $a_1<0$, which follows automatically from \eqref{H_positivity}, since $a_1-a_2<0$ and $a_2<0$ imply $a_1<a_2<0$. Moreover, discarding once more the zero mode along $x^1$, \eqref{eomPhi=0} becomes
\begin{equation}
    \left(\partial_0 + \frac{a_1}{2\kappa}\,\partial_1\right)\psi = 0\,,
\end{equation}
yielding a reduced chiral theory of Floreanini-Jackiw type \cite{Floreanini:1987as}. Because imposing $\phi=0$ directly in the action is a genuine phase-space reduction, its Dirac bracket is not obtained by simply substituting $\phi=0$ into the unreduced algebra \eqref{comm_rule}. In fact, the symplectic term in \eqref{S_phi0} gives
\begin{equation}
 [\partial_1\psi(X),\partial_1\psi(X')]_{\rm red}
 =\frac{i}{4\kappa}\,\partial_1\delta(x^1-x'^1)\,,
 \label{reduced_psi_bracket}
\end{equation}
which is the Dirac bracket of the reduced theory. This is consistent with treating the two realisations as distinct phase-space constructions.

The two realisations are complementary and coincide at the special point $a_1=2a_2$, where their respective chiral velocities also agree. The condition $\phi=0$ therefore has two distinct interpretations depending on how it is imposed: on-shell, it selects a chiral subsector of the full non-chiral theory \eqref{S_2D_psi}; at the level of the action, it defines a separate reduced chiral theory. These are not equivalent in general, and this distinction is the key to the comparison with the literature.

The first case, namely the on-shell realisation, is the natural one for making contact with the analysis of~\cite{Parasar:2025ngk}, where an emergent chiral boundary sector is found. The boundary spectrum there is richer, comprising both a chiral mode with linear dispersion and a chiral fractonic boson with dispersion $\omega\propto q^3$. What matters for our purposes is that the symmetry structure of the problem naturally allows a chiral boundary sector. In our framework the on-shell realisation exhibits such a mode as a consistent subsector of the full induced boundary dynamics, without postulating it at the level of the reduced action. The separate question of bulk--boundary matching is treated in Section~\ref{section_BB}.

The second case, namely the realisation at the level of the action, instead, makes the closest contact with Ref.~\cite{Prem:2017kxc}, where the boundary action is described by a scalar field entering through the combinations $\partial_0\partial_1\varphi$ and $\partial_1^2\varphi$, yielding a functional structure of the form
\begin{equation}
    \partial_0\partial_1\varphi\,\partial_1^2\varphi - v\,(\partial_1^2\varphi)^2\,.
\end{equation}
Indeed, \eqref{S_phi0} reproduces the same structure, with $\xi$ playing the role of $\varphi$:
\begin{equation}
    \int d^2X\left[\partial_0\partial_1\xi\,\partial_1^2\xi - v\,(\partial_1^2\xi)^2\right]\,,
\end{equation}
exhibiting a common intermediate description of the boundary in terms of a scalar potential entering only through higher-order derivatives. An analogous structure also appears in Ref.~\cite{Parasar:2025ngk}. In our framework, this scalar-potential representation is completed by identifying the dynamical field as $\psi=\partial_1\xi$, with $\phi$ auxiliary. After $\phi$ is integrated out, the boundary theory reduces to \eqref{S_2D_psi}, a Gaussian second-order theory for $\psi$, with a Luttinger sector on the positive-energy branch \eqref{H_positivity}. On this branch $a_2(a_1-a_2)>0$, and therefore the Luttinger velocity in \eqref{vk_id} is real. The structural contact with \cite{Prem:2017kxc} concerns the scalar-potential parametrisation, while the final boundary dynamics is encoded in $\psi$.

In summary, the on-shell realisation most clearly exhibits a chiral subsector of the full induced boundary dynamics and is conceptually most effective for the comparison with~\cite{Parasar:2025ngk}, while the realisation at the level of the action most transparently displays the structural connection with~\cite{Prem:2017kxc}. In either reading, the $h_{ij}$ sector admits a single propagating dynamical degree of freedom.

%------------------------------------------------------
\subsection{Physical information carried by the edge}
%------------------------------------------------------

The two boundary algebras determine the universal central/contact structure of the equal-time current algebra, without by themselves fixing the full correlators, while the Hamiltonian coefficients determine the propagation velocities. The
first derivative in the central term of \eqref{w_algebra} and the second derivative in \eqref{comm_rule} give distinct momentum dependence in the central/contact terms of the two equal-time current algebras. Together with the propagation generated by the boundary Hamiltonians, this distinction can appear in edge-current response and spectral functions and may provide an experimental discriminator of the higher-rank gauge structure.

The coexistence of the two boundary sectors also has a qualitative counterpart in the fractonic edge spectra of Refs.~\cite{Prem:2017kxc,Parasar:2025ngk}. In Ref.~\cite{Prem:2017kxc}, the higher-rank Chern--Simons edge supports two
co-propagating chiral modes associated with dipoles oriented perpendicular and parallel to the boundary: the former behaves as an ordinary chiral mode, while the latter retains the higher-rank constraint. Ref.~\cite{Parasar:2025ngk}
likewise separates a fractonic edge sector, with immobile charges and longitudinal dipoles but mobile quadrupoles and higher multipoles, from a non-fractonic chiral mode carried by transverse dipoles. In our covariant decomposition, the tangential $h_{ij}$ sector carries the fractonic rank-two current, whereas the normal--tangential field $w_i=h_{2i}$ carries the ordinarily conserved chiral current. This parallel is qualitative: the present boundary field theory does not by itself identify the propagating modes with specific microscopic dipolar or quadrupolar bound states.

The action \eqref{S_w} is the standard chiral-scalar action. On the positive-energy branch, after eliminating $\phi$, Eq.~\eqref{S_2D_psi} is a free Gaussian scalar action after a constant rescaling of time; on the $\phi=0$ branch it reduces to a second chiral scalar. The local edge dynamics can therefore be analysed with the methods of $(1+1)$-dimensional conformal field theory. Global information, including compactification and the spectrum of non-local operators, is fixed by the boundary conditions and by the microscopic realisation. These actions also provide the local edge input for entanglement calculations. At the level of the local critical edge theory, the corresponding entanglement contributions
can be analysed with the same $(1+1)$-dimensional conformal methods used for Chern-Simons edge modes. This analogy does not extend to the complete bulk-edge system: unlike in a purely topological Chern--Simons bulk, the edge
data alone do not determine the full bulk--edge entanglement structure. The finite-distance relation \eqref{mixed_bulk_boundary} involves gauge-fixed components, and the corresponding gauge-invariant response is analysed in Section~\ref{section_micro}.

%------------------------------------------------------
\subsection{Current-algebra symmetry of the \texorpdfstring{$w_i$}{wi} boundary sector}
%------------------------------------------------------

The $w_i$ current algebra provides a direct intrinsic characterization of the boundary transformation, without requiring a globally admissible affine bulk gauge transformation. On this branch, let $\alpha(X)$ satisfy the same chiral equation as the boundary current,
\begin{equation}
 (\partial_0+v_w\partial_1)\alpha=0\,.
 \label{alpha_chiral}
\end{equation}
Define the weighted, or smeared, charge:
\begin{equation}
 Q_w[\alpha](x^0)\equiv\int dx^1\,\alpha(X)w_1(X)\,.
 \label{Qw_def}
\end{equation}
Here ``smearing'' means multiplying the local current density $w_1$ by the weight function $\alpha$ and integrating over the spatial boundary coordinate. Using $w_1=\partial_1\chi$, Eq.~\eqref{EoM_chi} becomes $(\partial_0+v_w\partial_1)w_1=0$. Together with \eqref{alpha_chiral}, this yields
\begin{equation}
 \frac{d}{dx^0}Q_w[\alpha]
 =-v_w\int dx^1\,\partial_1(\alpha w_1)=0\,.
 \label{Qw_conservation}
\end{equation}
for the same endpoint conditions used in the current algebra. Equation \eqref{w_algebra} then gives
\begin{equation}
 [Q_w[\alpha],Q_w[\beta]]
 =\frac{i}{2\kappa}\int dx^1\,\alpha\,\partial_1\beta\,,
 \label{Qw_algebra}
\end{equation}
up to endpoint terms. With $G_w[\alpha]\equiv2\kappa Q_w[\alpha]$ one has
\begin{equation}
 i[G_w[\alpha],w_1(x^1)]=\partial_1\alpha(x^1)\,.
 \label{Qw_generator}
\end{equation}
Thus the charges $Q_w[\alpha]$ are conserved, their algebra \eqref{Qw_algebra} carries a central extension, and $G_w[\alpha]$ is the corresponding normalized canonical generator acting non-trivially on the physical boundary current. Locally, modulo the same zero mode used in $w_i=\partial_i\chi$, its action can be represented as
\begin{equation}
 \delta_\alpha\chi=\alpha\,,\qquad
 \delta_\alpha w_i=\partial_i\alpha\,.
 \label{chi_boundary_symmetry}
\end{equation}
This is a current-algebra symmetry of the chiral boundary phase space, not a second bulk gauge symmetry. The coefficient $\kappa$ in \eqref{S_w} is precisely the coefficient required to reproduce the central term in \eqref{w_algebra}, giving the same local Floreanini--Jackiw/Kac--Moody mechanism as for an Abelian Chern--Simons edge. The correspondence is intrinsic to the boundary sector. The complete bulk theory remains non-topological and carries
local degrees of freedom; the physical content of the finite-distance commutator~\eqref{mixed_bulk_boundary} is analysed in Section~\ref{section_micro}. If an admissible residual bulk transformation exists and its first normal boundary datum $\lambda_1$ satisfies the same chiral equation as $\alpha$, one may identify $\alpha=\lambda_1$. No such bulk extension is required to define the boundary current algebra or its canonical action.

%------------------------------------------------------
\subsection{Redundancy of the scalar parametrisation of the \texorpdfstring{$h_{ij}$}{hij} sector}
%------------------------------------------------------

Equation~\eqref{trasf_xi_phi} is not a residual bulk gauge transformation: it leaves $h_{ij}$ unchanged and expresses the non-uniqueness of the scalar parametrisation \eqref{defhx2=0}. It acts on the convenient local variable $\psi=\partial_1\xi$ as
\begin{equation}
    \delta\psi = b_1 - cX^1\,.
\end{equation}
The parameter $c$ also shifts $q=2\kappa\psi$, while the canonical momentum $p=\partial_1^2\xi+\phi$ is invariant: $\delta p=-c+c=0$. Thus global or asymptotic conditions fix the representative of this parametrisation redundancy. For the physical configuration space used here, it is natural to impose the asymptotic condition
\begin{equation}
    \psi \to 0 \qquad \text{as} \qquad |X^1|\to\infty\,,
\end{equation}
whose preservation requires $b_1=c=0$ and hence $\delta\psi = 0$. The parametrisation redundancy \eqref{trasf_xi_phi} is thus reduced, after imposing this asymptotic condition, to the subsector
\begin{equation}
    \delta\xi = a + b_0 X^0\,,\qquad \delta\phi = 0\,,
\end{equation}
which is compatible with the asymptotic condition $\psi\to 0$ defining the physical configuration space. In particular, this includes the constant shift of $\xi$, in analogy with standard boundary field theories \cite{Maggiore:2017vjf, Maggiore:2018bxr}.

%**********************************************
\section{Bulk-boundary correspondence}\label{section_BB}
%**********************************************

The two integrated boundary identities determine complementary intrinsic spatial current algebras; their mixed spatial commutator vanishes when both insertions lie on the boundary. We now match both sectors to the boundary conditions obtained from the bulk variational principle. For the explicit dynamical matching we work on the decoupled branch $c^{ijk}=0$, for which the induced action is \eqref{complete_boundary_action} and the $w_i$ and $h_{ij}$ boundary conditions are independent.

\subsection{The \texorpdfstring{$w_i$}{wi} sector}

Using $w_i=\partial_i\chi$, the two components of \eqref{BC_w} become
\begin{align}
 2c^{00}\,\partial_0\chi+(2c^{01}-\kappa)\partial_1\chi&=0\,,
 \label{BC_w0}\\
 (2c^{01}+\kappa)\partial_0\chi+2c^{11}\partial_1\chi&=0\,.
 \label{BC_w1}
\end{align}
A non-trivial boundary field exists when
\begin{equation}
 4\left[c^{00}c^{11}-(c^{01})^2\right]+\kappa^2=0\,.
 \label{w_det}
\end{equation}
Equation \eqref{w_det} is the compatibility condition among the boundary coefficients that selects the current-carrying $w_i$ branch. It plays the same role as \eqref{c5} in the $h_{ij}$ sector, where consistency among the three boundary conditions fixes one boundary coefficient in terms of the others. On the non-degenerate branch carrying the local current $w_1$, the two equations coincide and give
\begin{equation}
 \left(\partial_0+v_w\partial_1\right)\chi=0\,,\qquad
 v_w=\frac{2c^{01}-\kappa}{2c^{00}}
 =\frac{2c^{11}}{2c^{01}+\kappa}\,.
 \label{w_chirality}
\end{equation}
The first or second expression is used according to which denominator is non-zero; the degenerate null vector with $w_1=0$ contains only the spatial zero mode and carries no local current excitation. The equation of motion \eqref{EoM_chi} agrees with \eqref{w_chirality} for
\begin{equation}
 a_w=\kappa v_w\,.
 \label{aw_match}
\end{equation}
The boundary condition therefore fixes the chiral velocity and the Hamiltonian coefficient. The positive-energy condition $a_w<0$ becomes $\kappa v_w<0$.

\subsection{The \texorpdfstring{$h_{ij}$}{hij} sector}

In Section~\ref{section_2D} we showed that $\phi$ is auxiliary and can be integrated out, leaving $\psi$ as the unique propagating degree of freedom in the $h_{ij}$ sector. We now determine the branch selected by the $h_{ij}$ boundary conditions. The full induced action \eqref{S_2D} remains well defined with $\phi$ present, while the compatibility analysis identifies $\phi=0$ as the sector connected to the positive-energy branch.

The BC for $h_{ij}$ are obtained by substituting the boundary parametrisation \eqref{defhx2=0}, $h_{ij}|_{x^2=0}=\partial_i\partial_j\xi+\eta_{ij}\phi$. Introducing the compact notation
\begin{equation}
c_0\equiv c^{0000}\,,\quad c_1\equiv c^{0001}\,,\quad c_2\equiv c^{0011}\,,\quad c_3\equiv c^{0111}\,,\quad c_4\equiv c^{1111}\,,\quad c_5\equiv c^{0101}\,,
\label{c_notation}
\end{equation}
the three independent components of \eqref{BC_h} read
\begin{align}
2c_0\,\partial_0^2\xi + 2c_2\,\partial_1^2\xi + (4c_1+\kappa)\,\partial_0\partial_1\xi + 2(c_2-c_0)\phi &= 0\,,\label{BC00}\\
\bigl(2c_1-\tfrac{\kappa}{2}\bigr)\partial_0^2\xi + \bigl(2c_3-\tfrac{\kappa}{2}\bigr)\partial_1^2\xi + 4c_5\,\partial_0\partial_1\xi + 2(c_3-c_1)\phi &= 0\,,\label{BC01}\\
2c_2\,\partial_0^2\xi + 2c_4\,\partial_1^2\xi + (4c_3+\kappa)\,\partial_0\partial_1\xi + 2(c_4-c_2)\phi &= 0\,.\label{BC11}
\end{align}
The equations of motion obtained from $S_h$ \eqref{S_2D} by varying with respect to $\phi$ and $\xi$ are, respectively,
\begin{align}
2\kappa\,\partial_0\partial_1\xi + 2a_2(\partial_1^2\xi+\phi) &= 0\,,\label{EoMphiN}\\
2\kappa\,\partial_0\partial_1^3\xi + \kappa\,\partial_0\partial_1\phi + a_1\,\partial_1^4\xi + a_2\,\partial_1^2\phi &= 0\,.\label{EoMxiN}
\end{align}

Consider first the general case $\phi\neq 0$. We restrict to the non-degenerate branch $c_0\neq0$ and $c_2^2-c_0c_4\neq0$. Equation~\eqref{BC00} can then be used to eliminate $\partial_0^2\xi$. Imposing the consistency between \eqref{BC01} and \eqref{BC11} via
\begin{equation}
c_5 = \frac{\tfrac{\kappa^2}{16}(c_0-2c_2+c_4) + (2c_1c_2c_3 - c_0c_3^2 - c_1^2c_4)}{c_2^2-c_0c_4}\,,\label{c5}
\end{equation}
the two equations collapse to the same structure
\begin{equation}
(\partial_1^2\xi+\phi) + \Lambda_h\,\partial_0\partial_1\xi = 0\,,\qquad
\Lambda_h \equiv \frac{c_2(\kappa+4c_1) - c_0(\kappa+4c_3)}{2(c_2^2-c_0c_4)}\,.\label{BC_star}
\end{equation}
Comparison with \eqref{EoMphiN} immediately fixes
\begin{equation}
a_2 = \frac{\kappa}{\Lambda_h}\,.\label{a2match}
\end{equation}
Eliminating $\phi$ on-shell between \eqref{EoMphiN} and \eqref{EoMxiN} yields the wave equation
\begin{equation}
\Bigl[(a_1-a_2)\,\partial_1^2 - \frac{\kappa^2}{a_2}\,\partial_0^2\Bigr]\partial_1^2\xi = 0\,,\label{wave_general}
\end{equation}
with phase velocity $v^2 = a_2(a_1-a_2)/\kappa^2$. Under the strong matching criterion adopted here, namely that the boundary conditions imply the boundary equations of motion for the non-trivial local sector, compatibility between \eqref{wave_general} and the BC \eqref{BC_star} and \eqref{BC00} requires the coefficients of the independent differential operators to vanish after $\partial_0^2\xi$ has been eliminated through \eqref{BC00}. This gives two independent algebraic conditions. One of them is
\begin{equation}
a_2(a_1-a_2) = -\kappa^2 < 0\,,\label{nogo}
\end{equation}
which by itself conflicts with the Hamiltonian positivity condition \eqref{H_positivity}, since the latter implies $a_2(a_1-a_2)>0$. The other independent compatibility condition constrains only the boundary coefficients and is not needed for this conclusion. Thus the necessary condition \eqref{nogo} is already sufficient to obstruct positive-energy bulk-boundary matching in the $\phi\neq0$ sector. It arises because $\phi$, while auxiliary in the induced boundary theory, enters the bulk compatibility conditions before being eliminated. This selects $\phi=0$ as the sector in which bulk-boundary compatibility with the positive-energy branch \eqref{H_positivity} is realised.
For $\phi=0$, on the same non-degenerate branch, the BC \eqref{BC00}--\eqref{BC11} simplify. Using \eqref{BC00} again to eliminate $\partial_0^2\xi$ and imposing the consistency between \eqref{BC01} and \eqref{BC11} via \eqref{c5}, both reduce to the chiral equation
\begin{equation}
\left(\partial_0 + \frac{1}{\Lambda_h}\,\partial_1\right)\partial_1\xi = 0\,,\label{BCphi0}
\end{equation}
with $\Lambda_h$ given in \eqref{BC_star}. The matching is performed entirely within the full boundary theory \eqref{S_2D}, before imposing any reduced action. Setting $\phi=0$ in the full auxiliary-field equation \eqref{EoMphiN} gives
\begin{equation}
\left(\partial_0+\frac{a_2}{\kappa}\,\partial_1\right)\partial_1\xi=0\,.
\label{EoMphi0_full}
\end{equation}
Comparison with \eqref{BCphi0} fixes
\begin{equation}
a_2=\frac{\kappa}{\Lambda_h}\,.
\label{a2match_phi0}
\end{equation}
With $\phi=0$, the second full equation \eqref{EoMxiN}, together with \eqref{EoMphi0_full}, reduces for non-zero spatial momentum to
\begin{equation}
(a_1-2a_2)\,\partial_1^4\xi=0\,,
\end{equation}
so controlled compatibility of the non-trivial chiral sector requires
\begin{equation}
a_1=2a_2=\frac{2\kappa}{\Lambda_h}
=4\kappa\,\frac{c_2^2-c_0c_4}{c_2(\kappa+4c_1) - c_0(\kappa+4c_3)}\,.
\label{a1match}
\end{equation}
Thus the bulk boundary condition and both equations of motion of the full induced theory agree on the matched chiral subsector. This is precisely the point $a_1=2a_2$ at which the on-shell chiral subsector and the separately reduced Floreanini--Jackiw description discussed in Section~\ref{section_2D} have the same velocity; the reduced theory is not used to establish the matching.

On the positive-energy branch, \eqref{H_positivity} gives $a_2<0$ and $a_1-a_2=a_2<0$, while the $w_i$ sector requires $a_w<0$. Since $v_h=a_2/\kappa=1/\Lambda_h$ and $v_w=a_w/\kappa$, the two matched chiral modes obey
\begin{equation}
 \operatorname{sgn}(v_h)=\operatorname{sgn}(v_w)
 =-\operatorname{sgn}(\kappa)\,.
 \label{copropagation}
\end{equation}
They therefore co-propagate along the boundary. Their common orientation, together with the distinct first- and second-derivative current algebras, is a characteristic low-energy signature of the higher-rank edge theory.

%**********************************************
\section{Microcausality and gauge-invariant bulk response}\label{section_micro}
%**********************************************

\subsection{Gauge-invariant observables and equal-time cancellation}

The mixed commutator \eqref{mixed_bulk_boundary} is non-zero at equal-time spacelike separation and grows linearly with the normal distance. In this section we determine its physical content. The test must be performed on the local gauge-invariant bulk observables. For the rank-two gauge field $h_{\mu\nu}$, we use the scalar-gauge-invariant field strength introduced for the covariant bulk theory in Ref.~\cite{Bertolini:2024yur},
\begin{equation}
 F_{\mu\nu\rho}
 =\partial_\mu h_{\nu\rho}+\partial_\nu h_{\mu\rho}
 -2\partial_\rho h_{\mu\nu}\,,
 \label{field_strength_F}
\end{equation}
which satisfies $\delta F_{\mu\nu\rho}=0$ under \eqref{fracton_gauge}. This field strength is locally complete. Indeed, $F_{\mu\nu\rho}-F_{\mu\rho\nu}=3(\partial_\nu h_{\mu\rho}-\partial_\rho h_{\mu\nu})$; hence $F=0$ implies, by the Poincar\'e lemma and the symmetry of $h_{\mu\nu}$, that locally $h_{\mu\nu}=\partial_\mu\partial_\nu\lambda$. Two configurations with the same field strength therefore differ locally by a gauge transformation, so no local gauge-invariant quantity can distinguish them: the field strength and its derivatives generate the local gauge-invariant bulk observables, as in Maxwell and Yang--Mills theories \cite{Piguet:1995er}. Showing that a commutator vanishes on $F_{\mu\nu\rho}$ therefore shows that it vanishes on every such observable, and leaves none of them outside the test. The axial condition $h_{22}=0$ does not make the remaining components gauge invariant: residual transformations preserving the axial gauge satisfy $\partial_2^2\lambda=0$, and Eq.~\eqref{residual_lambda} gives $\delta_\lambda h_{11}=\partial_1^2\lambda_0+x^2\partial_1^2\lambda_1$. Physical microcausality must therefore be tested on the gauge-invariant field strength. A further reason is that Eq.~\eqref{mixed_bulk_boundary} refers to components defined relative to the axial gauge-fixing $h_{22}=0$, so the algebra of gauge-fixed components is a property of that choice and carries no meaning independent of it, whereas $F_{\mu\nu\rho}$ is invariant under \eqref{fracton_gauge} and is the same operator for any gauge fixing. The same distinction is familiar in electrodynamics in the Coulomb gauge: the scalar potential responds instantaneously to the charge density, but its apparently acausal contribution to the electric field is cancelled by the corresponding contribution from the vector potential, so that the local gauge-invariant
electromagnetic fields remain causal \cite{Jackson2002}.

Keeping the second insertion at finite $x^{\prime 2}>0$ in Eq.~\eqref{der_WI_w}, instead of taking its boundary limit, gives for $m=1$ the bulk--boundary $w_1$--$w_1$ commutator 
\begin{equation}
 \bigl[w_1(X),w_1(x')\bigr]_{x^0=x'^0}
 =\frac{i}{2\kappa}\,\partial_1\delta(x^1-x'^1)\,,\qquad x^{\prime2}>0.
 \label{w_bulk_boundary}
\end{equation}
Both this relation and Eq.~\eqref{mixed_bulk_boundary} descend from Eq.~\eqref{WI_w}, the first normal moment of the local bulk relation, obtained by weighting it with the normal coordinate before integrating over it. In that identity the source $J^1$ enters with unit weight in the normal coordinate and $J^{11}$ with weight $x^2$, so the corresponding functional derivatives produce an $x^{\prime2}$-independent and an $x^{\prime2}$-linear commutator respectively. The two are combined when a gauge-invariant field strength is formed.

The component of \eqref{field_strength_F} built from the spatial fields entering Eq.~\eqref{mixed_bulk_boundary} is
\begin{equation}
 F_{121}=\partial_2 h_{11}-\partial_1 h_{12}
 =\partial_2 h_{11}-\partial_1 w_1\,.
 \label{F121}
\end{equation}
Using Eqs.~\eqref{mixed_bulk_boundary} and \eqref{w_bulk_boundary}, one finds at equal times, in the distributional sense,
\begin{align}
 \bigl[w_1(X),F_{121}(x')\bigr]_{x^0=x'^0}
 &=\partial'_2\bigl[w_1(X),h_{11}(x')\bigr]
   -\partial'_1\bigl[w_1(X),w_1(x')\bigr]\nonumber\\
 &=-\frac{i}{2\kappa}\partial_1^2\delta(x^1-x'^1)
   +\frac{i}{2\kappa}\partial_1^2\delta(x^1-x'^1)=0\,.
 \label{F121_cancel}
\end{align}
The relative sign follows from $\partial'_1\delta(x^1-x'^1)=-\partial_1\delta(x^1-x'^1)$. The other non-trivial purely spatial component vanishes as well: in the axial gauge $F_{122}=\partial'_1h_{22}-\partial'_2h_{12}=-\partial'_2w_1$, so that
$\bigl[w_1(X),F_{122}(x')\bigr]=0$ follows directly from the $x^{\prime2}$-independence of Eq.~\eqref{w_bulk_boundary}. Equations~\eqref{mixed_bulk_boundary}, \eqref{w_bulk_boundary} and the axial condition $h_{22}=0$ can be collected as
\begin{equation}
 \begin{aligned}
 \bigl[w_1(X),h_{11}(x')\bigr]_{x^0=x'^0}
 &=-\frac{i}{2\kappa}\partial_1'^2
   \left[x^{\prime2}\delta(x^1-x'^1)\right],\\
 \bigl[w_1(X),h_{12}(x')\bigr]_{x^0=x'^0}
 &=-\frac{i}{2\kappa}\partial'_1\partial'_2
   \left[x^{\prime2}\delta(x^1-x'^1)\right],\\
 \bigl[w_1(X),h_{22}(x')\bigr]_{x^0=x'^0}
 &=-\frac{i}{2\kappa}\partial_2'^2
   \left[x^{\prime2}\delta(x^1-x'^1)\right]=0\,.
 \end{aligned}
 \label{spatial_Hessian_kernel}
\end{equation}
These three entries are the second derivatives of one and the same scalar function $x^{\prime2}\delta(x^1-x'^1)$, that is, its Hessian matrix, and we accordingly refer to the mixed commutator as a Hessian kernel. Because mixed partial derivatives commute, the differential operator defining the field strength annihilates any kernel of this form. Hence every spatial field-strength operator annihilates this equal-time kernel: all purely spatial field-strength components commute with the boundary current at equal times. This cancellation uses only the spatial commutators and the on-shell ordinary conservation law $\epsilon^{ij}\partial_iw_j|_{x^2=0}=0$ in Eq.~\eqref{on_shell}, which is the boundary operator constraint entering the derivation of Eq.~\eqref{w_bulk_boundary}. It requires neither the specialisation $c^{ijk}=0$ nor the chiral evolution used below.

\subsection{Unequal-time reconstruction}

For the unequal-time response we specialise, as in the explicit boundary dynamics of Sections~6 and 7, to the decoupled branch $c^{ijk}=0$. On this branch Eq.~\eqref{EoM_chi}, with $w_1=\partial_1\chi$, gives 
\begin{equation}
 (\partial_0+v_w\partial_1)w_1=0\,,\qquad v_w=\frac{a_w}{\kappa}\,.
 \label{w_chiral_current}
\end{equation}
We work in the local sector with non-zero momentum along the boundary, in which independent modes constant in the boundary spatial coordinate are excluded, as in the boundary construction of Section~\ref{section_2D}. Define
\begin{equation}
 r=x^1-x'^1-v_w(x^0-x'^0)\,,\qquad
 K_{\mu\nu}(X;x')\equiv\bigl[w_1(X),h_{\mu\nu}(x')\bigr] .
 \label{mixed_kernel_def}
\end{equation}
We call $K_{\mu\nu}$ the mixed kernel because it is the distributional two-point commutator between the boundary current and the bulk gauge field; smearing it in the boundary point $X$ gives the commutator of the corresponding
smeared boundary current with $h_{\mu\nu}(x')$. Equation~\eqref{w_chiral_current} holds as an operator equation, so applying $(\partial_0+v_w\partial_1)$ to the boundary argument of $K_{\mu\nu}$ gives $(\partial_0+v_w\partial_1)K_{\mu\nu}=0$. Only the unordered commutator is involved here: no time-ordered product is differentiated and no contact term is generated. Consequently $K_{\mu\nu}$ depends on the boundary coordinates
only through $x^1-v_wx^0$ and, by time-translation invariance, only through $r$. Matching at $x^0=x'^0$ to the equal-time data~\eqref{spatial_Hessian_kernel} then gives
\begin{equation}
 K_{11}=-\frac{i x^{\prime2}}{2\kappa}\delta''(r)\,,\qquad
 K_{12}=\frac{i}{2\kappa}\delta'(r)\,,\qquad
 K_{22}=0\,.
 \label{mixed_kernel_spatial}
\end{equation}
For $x^{\prime2}>0$, the source-free bulk parts of Eqs.~\eqref{EoM_w}--\eqref{EoM_hij}, in the axial gauge $h_{22}=0$, imply
\begin{align}
 \partial'_2 h_{01}&=\partial'_1 w_0,
 &\partial'_2 h_{01}&=\partial'_0 w_1,\nonumber\\
 \partial'_1 h_{01}&=\partial'_0 h_{11}+\partial'_2 w_0,
 &\partial'_1 h_{00}&=\partial'_0 h_{01}+\partial'_2 w_1,\nonumber\\
 \partial'_2(h_{00}+h_{11})&=\partial'_0 w_0+\partial'_1 w_1 .
 \label{bulk_component_relations}
\end{align}
Taking their unordered commutators with $w_1(X)$, together with
Eq.~\eqref{mixed_kernel_spatial}, fixes
\begin{equation}
 K_{02}=-v_wK_{12}\,,\qquad
 K_{01}=-v_wK_{11}\,,\qquad
 K_{00}=v_w^2K_{11}\,.
 \label{mixed_kernel_temporal}
\end{equation}
The temporal components are uniquely determined by the spatial ones in this local sector. Indeed, if $\Delta_{\mu\nu}$ is the difference of two kernels with the same spatial data, then $\Delta_{11}=\Delta_{12}=\Delta_{22}=0$. The homogeneous form of Eq.~\eqref{bulk_component_relations} gives successively
\begin{align}
 \partial'_2\Delta_{01}=\partial'_0\Delta_{12}=0,
 &\qquad \partial'_1\Delta_{02}=\partial'_2\Delta_{01}=0,\nonumber\\
 \partial'_1\Delta_{01}
 &=\partial'_0\Delta_{11}+\partial'_2\Delta_{02}=0,\nonumber\\
 \partial'_1\Delta_{00}
 &=\partial'_0\Delta_{01}+\partial'_2\Delta_{12}=0.
 \label{mixed_kernel_uniqueness}
\end{align}
A quantity independent of the tangential spatial coordinate is a spatial zero mode, and such modes do not belong to the local boundary phase space used above, which is built from $\partial_1\chi$. Excluding them, Eq.~\eqref{mixed_kernel_uniqueness} yields $\Delta_{02}=\Delta_{01}=\Delta_{00}=0$.

The full mixed kernel is consequently
\begin{equation}
 K_{\mu\nu}(X;x')
 =-\frac{i}{2\kappa}\partial'_\mu\partial'_\nu
 \left[x^{\prime2}\delta\!\left(r\right)\right].
 \label{mixed_kernel_Hessian}
\end{equation}
Its Hessian form is exactly the differential structure of the scalar gauge transformation~\eqref{fracton_gauge}: the non-local contribution lies in the scalar-gauge direction of the bulk gauge field, and the cancellation in the local field strength is structural rather than accidental. At every point the contribution can be absorbed by a local gauge transformation, so the boundary current shifts the gauge representative of the bulk configuration and not the configuration itself; by the local completeness of $F_{\mu\nu\rho}$, no local gauge-invariant bulk observable responds to it.

The reconstruction never imposed the boundary conditions of Section~\ref{section_BB}, yet it respects them. Equations~\eqref{BC_w0}--\eqref{BC_w1} are homogeneous linear operator relations among $w_0$ and $w_1$ at $x^2=0$; since the commutator is linear, $(K_{02},K_{12})$ at $x^{\prime2}=0$ must lie along their null vector. It does: the tangential entries $K_{ij}$ vanish there, while $(K_{02},K_{12})\propto(-v_w,1)$, which is that null vector at the matched velocity~\eqref{w_chirality}.

\subsection{Boundary generator and field-strength response}

The Hessian structure has a direct current-algebra interpretation. Using the mixed kernel~\eqref{mixed_kernel_Hessian} to evaluate the action of the conserved boundary generator $G_w[\alpha]=2\kappa Q_w[\alpha]$ of Eqs.~\eqref{Qw_def} and \eqref{Qw_generator}, one finds
\begin{equation}
 i\bigl[G_w[\alpha],h_{\mu\nu}(x')\bigr]
 =\partial'_\mu\partial'_\nu
 \left[x^{\prime2}\alpha(X')\right].
 \label{Qw_bulk_action}
\end{equation}
The parameter $\alpha$ obeys the chiral equation~\eqref{alpha_chiral} that ensures conservation of $G_w[\alpha]$, and the same condition preserves the chiral boundary equation~\eqref{w_chirality} under the local identification
$\alpha=\lambda_1$ discussed below Eq.~\eqref{chi_boundary_symmetry}. Locally, $\lambda_{\rm eff}=x^{\prime2}\alpha(X')$ obeys $\partial_2'^2\lambda_{\rm eff}=0$, with boundary data $\lambda_0=0$ and
$\lambda_1=\alpha$: the action of the boundary Kac--Moody generator on local bulk fields has precisely the differential form of the gauge-preserving affine profile~\eqref{residual_lambda}. The dependence on the boundary time drops out for the same reason: the smearing integral gives $\alpha$ evaluated at $x'^1+v_w(x^0-x'^0)$, which equals $\alpha(X')$ precisely when $\alpha$ is chiral.

This is the standard Chern--Simons boundary mechanism: gauge transformations that act non-trivially at the boundary become physical symmetries with non-zero charges and a central extension \cite{Witten:1988hf,Elitzur:1989nr}. Equation~\eqref{Qw_bulk_action} makes explicit how this boundary mechanism acts on the bulk fields of the present model: the boundary generator acts non-trivially on gauge-dependent bulk components along the local residual gauge direction, while the local gauge-invariant field strength is insensitive to that action. Indeed, applying the differential operator in Eq.~\eqref{field_strength_F} gives
\begin{align}
 \bigl[w_1(X),F_{\mu\nu\rho}(x')\bigr]
 &=\partial'_\mu K_{\nu\rho}+\partial'_\nu K_{\mu\rho}
   -2\partial'_\rho K_{\mu\nu}=0\,.
 \label{field_strength_zero}
\end{align}
The traceless field strength $\bar F_{\mu\nu\rho}$ used in Ref.~\cite{Bertolini:2024yur} is an algebraic projection of $F_{\mu\nu\rho}$; hence
\begin{equation}
 \bigl[w_1(X),\bar F_{\mu\nu\rho}(x')\bigr]=0\,,\qquad x^{\prime2}>0.
 \label{traceless_field_strength_zero}
\end{equation}
The generalized electric and magnetic fields of the covariant bulk theory are linear combinations of $\bar F$, so the same conclusion applies to them. Correspondingly, for a weak perturbation coupled to the boundary current, the retarded response in this channel is
\begin{equation}
 G^R_{\bar F w_1}(x';X)
 =-i\,\theta(x'^0-x^0)
 \left\langle\bigl[\bar F_{\mu\nu\rho}(x'),w_1(X)\bigr]\right\rangle=0\,,
 \qquad x^{\prime2}>0.
 \label{retarded_field_response}
\end{equation}
Here the field-strength operator has been applied to the unordered commutator before the retarded step function is introduced.

At equal time, every bulk insertion with $x^{\prime2}>0$ is spacelike separated from the boundary insertion. The vanishing field-strength commutators therefore verify the equal-time microcausality condition in this mixed bulk-boundary channel without imposing $c^{ijk}=0$. On that decoupled branch, and within the local non-zero-spatial-momentum sector specified above, Eq.~\eqref{retarded_field_response} holds for every $x^{\prime2}>0$ and at all times. It therefore establishes microcausality in this mixed response channel and is stronger than microcausality alone requires. The bulk theory remains non-topological and carries local physical degrees of freedom.

\subsection{Independent gauge-invariant source check}

As an independent check, couple a smooth test source $f$ with compact support strictly in $x^{\prime2}>0$ to the gauge-invariant component $F_{121}=\partial'_2h_{11}-\partial'_1w_1$. The interior support makes the identification of the sources exact: it removes the boundary term produced by the integration by parts and generates no sources localised at $x^2=0$, so the probe is a purely bulk configuration coupled gauge-invariantly. The identity~\eqref{WI_w} holds for arbitrary sources; the present subfamily is chosen to ask whether a gauge-invariant bulk probe excites the boundary breaking term. Integration by parts gives $J^{11}=-\partial'_2f$ and $J^1=\partial'_1f$. Their contribution to Eq.~\eqref{WI_w} is
\begin{align}
 \int_0^\infty dx^{\prime2}
 \left(x^{\prime2}\partial_1'^2J^{11}-\partial'_1J^1\right)
 &=-\int_0^\infty dx^{\prime2}\,
 \partial'_2\left(x^{\prime2}\partial_1'^2f\right)=0\,.
 \label{gauge_invariant_source_check}
\end{align}
Thus the same cancellation follows directly from the linked source components of a gauge-invariant probe.

The mixed commutator~\eqref{mixed_bulk_boundary} therefore describes the action of the boundary current algebra on gauge-dependent bulk components along the local scalar-gauge direction. Its Hessian structure explains why this action disappears identically from the local gauge-invariant field strength. Since the field strength and its derivatives generate the local gauge-invariant bulk observables, the cancellation extends from $F_{\mu\nu\rho}$ to every such observable, the commutators involved being $c$-numbers. At equal times, and without restricting the boundary parameters, this covers the spatial field-strength components; on the decoupled branch $c^{ijk}=0$, and within the local non-zero-spatial-momentum sector considered here, it covers the entire local gauge-invariant bulk algebra at every $x^{\prime2}>0$ and at all times.

%**********************************************
\section{Non-planar boundaries}\label{section_curved}
%**********************************************

The planar construction admits a geometric formulation that displays the new terms generated by boundary curvature and makes the covariant structure of the residual transformations explicit. Let $n^\mu$ be the unit normal to a generic boundary, $e_a{}^\mu$ a tangential frame and $\gamma_{ab}$ the induced metric. We denote by $\nabla_\mu$ the bulk Levi-Civita covariant derivative and by $D_a$ the intrinsic Levi-Civita covariant derivative associated with $\gamma_{ab}$; the normal derivative is $\nabla_n\equiv n^\mu\nabla_\mu$. The extrinsic curvature is $K_{ab}=e_a{}^\mu e_b{}^\nu\nabla_\mu n_\nu$. The tensor field decomposes as
\begin{equation}
 \sigma=n^\mu n^\nu h_{\mu\nu}\,,\qquad
 w_a=n^\mu e_a{}^\nu h_{\mu\nu}\,,\qquad
 h_{ab}=e_a{}^\mu e_b{}^\nu h_{\mu\nu}\,.
 \label{geometric_decomp}
\end{equation}
Because the bulk spacetime is flat, the Cartesian gauge transformation $\delta h_{\mu\nu}=\partial_\mu\partial_\nu\lambda$ can be written in arbitrary coordinates as the covariant Hessian $\delta h_{\mu\nu} \nabla_\mu\nabla_\nu\lambda$. In Gaussian normal coordinates, the projected gauge transformations are 
\begin{align}
 \delta h_{ab}&=D_aD_b\lambda+K_{ab}\nabla_n\lambda\,,
 \label{curved_delta_h}\\
 \delta w_a&=D_a(\nabla_n\lambda)-K_a{}^bD_b\lambda\,.
 \label{curved_delta_w}
\end{align}
Defining the boundary value and first normal derivative
\begin{equation}
 \lambda_0\equiv\lambda|_{\partial M}\,,\qquad
 \lambda_1\equiv\nabla_n\lambda|_{\partial M}\,,
 \label{curved_lambda_data}
\end{equation}
their action on the projected boundary fields is
\begin{align}
 \delta h_{ab}|_{\partial M}&=D_aD_b\lambda_0+K_{ab}\lambda_1\,,\notag\\
 \delta w_a|_{\partial M}&=D_a\lambda_1-K_a{}^bD_b\lambda_0\,.
 \label{curved_boundary_transformations}
\end{align}
Thus $\lambda_0$ and $\lambda_1$ are precisely the boundary value and first normal derivative of the scalar gauge parameter. For a plane, $K_{ab}=0$, and these two boundary data reduce to those used in the planar analysis; the zeroth and first normal projections of the scalar functional relation then give \eqref{WI} and \eqref{WI_w}. For a curved boundary, the extrinsic-curvature terms mix the tangential and normal structures entering those projections. The boundary conditions, the integrated boundary identities and the induced Hamiltonian can consequently contain local terms built from $K_{ab}$ and its derivatives, producing a given-geometry dependence of the edge spectrum.

The field $\psi=\partial_1\xi$ used on the plane is a spatial tangential derivative. If $s^a$ denotes the unit spacelike tangent selected by a chosen foliation of a generic timelike $(1+1)$-dimensional boundary into spatial slices, its geometric counterpart is
\begin{equation}
 \psi=s^aD_a\xi\,.
 \label{psi_curved}
\end{equation}
The covariant replacement of $\partial_1\xi$ is therefore the tangential quantity in \eqref{psi_curved}. By contrast, a normal derivative $\nabla_n\xi$ would require an off-boundary extension of $\xi$ and represents separate normal data rather than the covariantisation of $\partial_1\xi$. Equations \eqref{curved_delta_h} and \eqref{curved_delta_w} identify the curvature couplings absent in the planar theory and provide the starting point for their complete local classification.

The $K_{ab}$-dependent terms carry information about the embedding of the boundary in the bulk, not only about its intrinsic metric. This goes beyond the standard intrinsic Luttinger description: here the extrinsic curvature enters already in the projected gauge transformations and can consequently induce embedding-dependent corrections to the boundary conditions, Ward identities and edge Hamiltonian. Corresponding corrections to the edge response or spectrum would provide a potential geometric discriminator relative to the standard intrinsic Luttinger limit and a characteristic imprint of the non-topological higher-rank bulk that is hidden in the planar geometry. A complete determination of these curvature-dependent corrections requires the full curved-boundary analysis and lies beyond the scope of the present work.

%**********************************************
\section{Conclusions}
%**********************************************

We have determined the planar edge structure of the covariant $(2+1)$-dimensional higher-rank Chern-Simons theory of fractons. The differential generator associated with the single scalar gauge symmetry singles out one scalar Ward combination of the functional field equations. Its zeroth and first normal moments produce the two integrated boundary identities relevant to the edge: the zeroth moment gives the traceless rank-two current $\mathcal J^{ij}$ with the  fractonic conservation law $\partial_i\partial_j\mathcal J^{ij}=0$, while the first moment retains the $w_i$ sector and gives the vector current $\mathcal K^i$ with the ordinary conservation law $\partial_i\mathcal K^i=0$. These two boundary-relevant projections are naturally associated with the boundary value and first normal derivative of the same scalar gauge parameter, without introducing a second bulk gauge symmetry.

The two identities determine complementary equal-time algebras. The tangential tensor field carries the higher-rank Kac-Moody algebra with a second spatial derivative in the central term. The normal-tangential field carries a standard
$U(1)$ Kac-Moody algebra with a first spatial derivative. The mixed commutator between the boundary current $w_1$ and the gauge-fixed bulk component $h_{11}$ is proportional to the normal distance and vanishes when the bulk insertion is brought to the boundary. At equal time, its projection onto the local gauge-invariant spatial field-strength components vanishes without any restriction on the boundary parameters. Since every finite normal separation is then
spacelike, this cancellation verifies the equal-time microcausality condition in the mixed bulk-boundary channel. On the decoupled branch $c^{ijk}=0$, and within the local non-zero-spatial-momentum sector, the full unequal-time mixed kernel is a pure Hessian and is annihilated by every field-strength component. The corresponding retarded response consequently vanishes at all times, establishing microcausality in the same mixed response channel. This all-time result is stronger than microcausality alone requires. On the decoupled branch $c^{ijk}=0$, the direct-sum induced action has a block-diagonal intrinsic boundary symplectic form.

The conserved currents lead locally to $w_i=\partial_i\chi$ and $h_{ij}=\partial_i\partial_j\xi+\eta_{ij}\phi$. On the decoupled branch $c^{ijk}=0$, the induced action is $S_{\partial}=S_w[\chi]+S_h[\xi,\phi]$. The $w_i$ sector is a chiral scalar;
its boundary condition fixes the velocity through \eqref{w_chirality} and the Hamiltonian coefficient through \eqref{aw_match}. In the $h_{ij}$ sector, $\phi$ is auxiliary and $\psi=\partial_1\xi$ is the propagating field. Eliminating $\phi$ gives a non-chiral Luttinger liquid on the positive-energy branch \eqref{H_positivity}. The bulk boundary conditions obstruct the positive-energy matching for $\phi\neq0$ and select the chiral $\phi=0$ sector. Matching \eqref{BCphi0} to the two equations of motion of the full induced theory fixes $a_2=\kappa/\Lambda_h$ and $a_1=2a_2$, so that $v_h=a_2/\kappa=1/\Lambda_h$. The positive-energy conditions imply \eqref{copropagation}: the two chiral modes co-propagate, with their common orientation fixed by the sign of $\kappa$.

The first- and second-derivative central terms have distinct momentum structures in the two equal-time edge-current algebras. Together with the propagation velocities, they provide characteristic low-energy field-theoretical signatures of the higher-rank boundary. The chiral and Gaussian bosonic actions also connect the edge theory with $(1+1)$-dimensional conformal methods and provide the local input for response and entanglement calculations. At the level of the local critical edge, the latter can be treated with the same conformal methods used for Chern-Simons edge modes. For the complete theory, however, the non-topological bulk carries local physical degrees of freedom, so the edge
CFT data alone need not determine the full bulk-edge entanglement structure.

For a non-planar boundary, the projected gauge transformations contain the extrinsic curvature as in \eqref{curved_delta_h} and \eqref{curved_delta_w}. Curvature can mix the tangential and normal boundary data and modify the boundary conditions, current identities and edge Hamiltonian. Since these terms depend on the embedding of the boundary and not only on its intrinsic metric, they can provide a geometric discriminator relative to the standard Luttinger limit. Their complete determination requires a separate curved-boundary analysis.

The resulting two-sector edge structure is also qualitatively consistent with Refs.~\cite{Prem:2017kxc,Parasar:2025ngk}, where fractonic and ordinary or non-fractonic chiral edge modes are associated with distinct dipolar or multipolar sectors. Our comparison is at the level of the boundary current and mode structure, not a microscopic identification of the excitations.

Further directions include the boundary analysis of the massive Maxwell-Chern-Simons extension \cite{Bertolini:2025goo}, covariant fractonic BF theory \cite{Bertolini:2025jul}, and related higher-rank Chern-Simons models \cite{Bertolini:2025qcy}. Coupling the boundary currents to matter would connect the edge response directly with the Hall-like dipole transport and the fracton-vortex duality of Ref.~\cite{Bertolini:2024yur}. The relation with higher-order topological insulators \cite{You:2019bvu} and fracton-elasticity duality \cite{Pretko:2017kvd} provides a natural route to microscopic realisations. It would also be interesting to clarify whether the higher-rank chiral boundary mode admits a more direct interpretation in terms of quadrupolar bound states, in closer analogy with the edge spectrum discussed in Ref.~\cite{Parasar:2025ngk}.

%**************************************************************
%\section*{Acknowledgments}
%**************************************************************

%********************************************************

%********************************************************

\end{document}